\documentclass[journal,draftcls,onecolumn,12pt,twoside]{IEEEtranTCOM}
\ifCLASSINFOpdf
\else
\fi
\usepackage{color}

\usepackage{algorithm} %format of the algorithm
\usepackage{algorithmic} %format of the algorithm
\usepackage{graphicx}
\usepackage{epstopdf}
\usepackage{graphicx}
\usepackage{float}
\usepackage{color}
\usepackage{amsmath}
\usepackage{amsfonts}
\usepackage{amssymb}
\usepackage{amsthm}
\usepackage{slashbox}
\usepackage{bbding}
\usepackage{cite}
\usepackage{graphicx}
\usepackage{dsfont}
\usepackage{clrscode}
\usepackage{epstopdf}
\usepackage[T1]{fontenc}
\usepackage{url}

\usepackage{setspace}

\DeclareMathOperator*{\maxmin}{max\,min}

\makeatletter

\newcommand{\Rmnum}[1]{\expandafter\@slowromancap\romannumeral #1@}
\makeatother

\DeclareMathOperator*{\argmax}{\arg\!\max}

\newtheorem{theorem}{Theorem}
\newtheorem{lemma}{Lemma}
\newtheorem{proposition}{Proposition}

\begin{document}
\title{Distributed Channel Quantization for Two-User Interference Networks}
\author{\IEEEauthorblockN{Xiaoyi Leo Liu, Erdem Koyuncu, and Hamid Jafarkhani}\\
\IEEEauthorblockA{
Center for Pervasive Communications and Computing \\University of California, Irvine}
}
\maketitle
\begin{abstract}
We introduce conferencing-based distributed channel quantizers for two-user interference networks where interference signals are treated as noise. Compared with the conventional distributed quantizers where each receiver  quantizes its own channel independently, the proposed quantizers allow multiple rounds of feedback communication in the form of conferencing between receivers. We take the network outage probabilities of sum rate and minimum rate as performance measures and consider quantizer design in the transmission strategies of time sharing and interference transmission. First, we propose distributed quantizers that achieve the optimal network outage probability of sum rate for both time sharing and interference transmission strategies with an average feedback rate of only two bits per channel state. Then, for the time sharing strategy, we propose a distributed quantizer that achieves the optimal network outage probability of minimum rate with finite average feedback rate; conventional quantizers require infinite rate to achieve the same performance. For the interference transmission strategy, a distributed quantizer that can approach the optimal network outage probability of minimum rate closely is also proposed. Numerical simulations confirm that our distributed quantizers based on conferencing outperform the conventional ones.
\end{abstract}
%\begin{IEEEkeywords}
%distributed quantizer, conferencing, network outage probability, time sharing, interference transmission
%\end{IEEEkeywords}
%\IEEEpeerreviewmaketitle

\section{Introduction}

Channel quantization in a network with multiple receivers is fundamentally different from that in a point-to-point system. In a point-to-point system, the receiver can acquire the entire channel state information (CSI) and send the corresponding quantized feedback information to the transmitter \cite{Quantization_Interference,ErdemRelayFeedback,BDRaoTransmitBeamforming,ErdemQuantization}. On the other hand, in a network with multiple receivers, each receiver only has access to its own local CSI due to different geographical locations of the different receivers. Each receiver can thus quantize only a part of the entire global CSI, which results in a distributed quantization problem.

In the existing work on distributed quantization for networks  \cite{Quantization_Interference,Interference_Power_Control, Interference_Throughput},  each receiver first quantizes its local CSI independently and then sends a finite number of bits representing quantized information through feedback links to other terminals. After decoding feedback information from all receivers, each terminal reconstructs the quantized version of the global CSI. Afterwards, transmission methods such as beamforming or power control are adopted by treating the global quantized CSI as the exact unquantized CSI. For example, power control and throughput maximization for interference networks based on separate quantized feedback information from receivers are analyzed in \cite{Interference_Power_Control,Interference_Throughput}. In \cite{Quantization_Interference}, beamformers are designed for the $K$-user MIMO interference channels with independent quantized information from each receiver. The performance of these quantizers depend on the number of feedback bits assigned for quantization to each receiver and always suffer from some loss when compared with the optimal performance.

In this paper, we propose a novel distributed quantization strategy with multiple rounds of feedback communication in the form of conferencing between receivers. Through conferencing among receivers, partial CSI from other receivers can be utilized for a better overall quantizer performance. To illustrate this, we consider the distributed quantization problem for two-user interference networks with time sharing and interference transmission strategies. The network outage probability is the performance metric. We first propose a distributed quantizer that achieves the optimal network outage probability of sum rate in both time sharing and interference transmission with only two bits of feedback information. We also propose a distributed quantizer that attains the optimal network outage probability of minimum rate in time sharing with finite average feedback rate. For the optimal network outage probability of minimum rate in interference transmission, a distributed quantizer that can approach it closely is also proposed. By numerical simulations, we show the effectiveness of the proposed quantizers by comparing them with the conventional ones.

The rest of this paper is organized as follows: In Section \ref{secprelim},
we provide a description of the system model. In Sections
\ref{secthree} and \ref{secfour}, we introduce and analyze the distributed quantizers for time sharing and interference transmission strategies, respectively. Numerical simulations are provided in Section \ref{secnumresults}.

	{\bf Notations:} Bold-face letters refer to vectors or matrices. $\top$ denotes the matrix transpose. $\mathtt{C}$, $\mathtt{R}$ and $\mathtt{N}$ represent the sets of complex, real and natural numbers, respectively. The set of complex $n$-vectors is denoted by $\mathtt{C}^{n\times 1}$ and the set of complex $m\times n$ matrices is denoted by $\mathtt{C}^{m\times n}$.  $\mathtt{CN}(a, b)$ represents a circulary-symmetric complex Gaussian random variable (r.v.) with mean $a$ and covariance $b$.   $f_{X}(\cdot)$ is the probability density function (PDF) of  a r.v. $X$. $\left|\mathcal{S}\right|$ is the cardinality of the set $\mathcal{S}$. For sets $\mathcal{A}$ and $\mathcal{B}$, $\mathcal{A} - \mathcal{B} = \left\{x \in \mathcal{A}, x \notin \mathcal{B}\right\}$. $\textmd{E}[\cdot]$ denotes the expectation and $\textmd{Prob}\{\cdot\}$ denotes the probability. For any $x \in \mathtt{R}$, $\lfloor x\rfloor$ is the largest integer that is less than or equal to $x$ and $\lceil x \rceil$ is the smallest integer that is larger than or equal to $x$. For any logical statement $\sf ST$, we let $\bf{1}({{\sf ST}})=1$ when ${\sf ST}$ is true, and $\bf{1}({{\sf ST}})=0$ when ${\sf ST}$ is false. Finally, for $b_1,\ldots,b_N\in\{0,1\},\,N\geq 1$, the real number $[0.b_1\cdots b_N]_2$ is the base-$2$ representation of the real number $\sum_{n=1}^{N} b_n2^{-n}$.

\section{Preliminaries}
\label{secprelim}
\subsection{System strategy}
Consider an interference network where transmitters ${\sf S}_1$ and ${\sf S}_2$ send independent signals to receivers ${\sf D}_1$ and ${\sf D}_2$ concurrently. Both transmitters and receivers are equipped with only a single antenna. The channel gain from ${\sf S}_k$ to ${\sf D}_l$ is denoted by $h_{k, l}$ for $k, l = 1, 2$.  We assume that $h_{1, 1}, h_{2, 2} \simeq \mathtt{CN}(0, 1)$ and $h_{1, 2}, h_{2, 2} \simeq \mathtt{CN}(0, \epsilon)$, where  $\epsilon $ is the covariance of interference links. Let ${H}_{k, l} = \left|h_{k, l}\right|^2$. Then, ${\bf h}_k = \left[ H_{1, k}, H_{2, k}\right]^{\top}\in \mathtt{C}^{2\times 1}$ denotes the local CSI at receiver $k$, and ${\bf H} = \left[ {\bf h}_1, {\bf h}_2\right]\in \mathtt{C}^{2\times 2}$ represents the entire CSI. The additive noises at the receivers are distributed as $\mathtt{CN}(0, 1)$.

We assume a quasi-static block fading channel in which the channels vary independently from one block to another while remain constant within each block. Each receiver can perfectly estimate its local CSI and provide quantized instantaneous CSI to other terminals via error-free and delay-free feedback links.

\subsection{Transmission strategies}
We consider two transmission strategies in the two-user interference network, namely time sharing and interference transmission. Time sharing means either transmitter only occupies a proportion of the block to transmit while remains silent in the rest, thus no interference exists. Interference transmission refers to the scenario where both transmitters send signals within the entire block, thereby causing interference to each other. We assume that interference signals are dealt with as noises. Since we focus on the design of distributed quantizers based on conferencing, we also assume that only one strategy will be performed in the entire transmission for simplicity.

 In time sharing, let $t_k \in [0, 1]$ be the percentage of time within the entire block in which only ${\sf S}_k$ is active for $k = 1, 2$ with $t_1 + t_2 = 1$. The instantaneous power used by ${\sf S}_k$ is $P_k = p_k P$, where $p_k \in [0, 1]$ and $P$ is the short-term power constraint. It is optimal for both transmitters to use full power under the condition of no interference. Therefore, for a given $\bf H$, the end-to-end rate at receiver $k$ is
\begin{align}
\textit{R}_{{\sf ts}, k} (t_k) \triangleq t_k \log_2\left(1 + P H_{k, k}\right).\nonumber
\end{align}
In interference transmission, for $k, l = 1, 2$ and $k \neq l$, the end-to-end rate at receiver $k$ is
\begin{align}
\textit{R}_{{\sf it}, k} (p_1, p_2) \triangleq \log_2\left(1 + \frac{p_k P H_{k, k}}{p_l P H_{l, k} + 1}\right).\nonumber
\end{align}

\subsection{Network Outage Probability}
Our performance measure is the network outage probability, which is the fraction of channel states at which the rate measure of the network falls below a target data rate $\rho$.  Such a performance metric is well-suited for applications where a given constant data rate needs to be sustained for every channel state. Two kinds of rate measurements are considered, namely sum rate and minimum rate.  Our goal is to design efficient distributed quantizers that can achieve the optimal network outage probability of sum rate or minimum rate for both time sharing and interference transmission strategies.

\section{Distributed Quantization for Network Outage Probability of Sum Rate}
\label{secthree}
We first design distributed quantizers for interference transmission. The sum rate is $\textit{SR}_{\sf it}\left(p_1, p_2\right) \triangleq \sum_{k = 1}^2 \textit{R}_{{\sf it}, k} (p_1, p_2)$. We define the network outage probability as\footnote{We choose the sum-rate outage threshold to be $2\rho$ for a more fair comparison with the rate threshold $\rho$ that we shall specify for the minimum-rate outage threshold.}
 \begin{align}
   \textmd{OUT}_{\sf it}^{\sf sr} \triangleq \textmd{Pr}
\left\{
\textit{SR}_{\sf it}\left(p_1, p_2\right) < 2\rho
\right\}.\nonumber
 \end{align}
 It is proved in \cite{OptimalPowerControlSumRate} that the maximum sum rate  is $\max\left\{\textit{SR}_{\sf it}\left(1, 0\right), \textit{SR}_{\sf it}\left(0, 1\right), \textit{SR}_{\sf it}\left(1, 1\right)\right\}$. Therefore, the optimal (minimum-achievable) network outage probability is
\begin{align}
\textmd{OUT}_{{\sf sr}, \sf it}^{ {\sf opt}}\! =\!
\textmd{Pr}\left\{
\max\left\{\textit{SR}_{\sf it}\left(1, 0\right), \textit{SR}_{\sf it}\left(0, 1\right), \textit{SR}_{\sf it}\left(1, 1\right)\right\} \!<\! 2\rho
\right\}.\nonumber
\end{align}

In the following, we design a distributed quantizer, namely ${\textmd{\textit{DQ}}}_{{{{\sf sr}, {\sf it}}}}$, that can achieve $\textmd{OUT}_{{\sf sr}, \sf it}^{ {\sf opt}}$ with only $1$ feedback bit per receiver. The quantizer ${\textmd{\textit{DQ}}}_{{{{\sf sr}, {\sf it}}}}$ consists of two local encoders and a unique decoder. The $k$-th encoder $\textmd{ENC}_{{\sf sr}, {\sf it}, k}$ is located at receiver $k$ and the decoder $\textmd{DEC}_{{\sf sr}, {\sf it}}$ is shared by all terminals, for $k = 1, 2$. The components of ${\textmd{\textit{DQ}}}_{{{{\sf sr}, {\sf it}}}}$  operate as follows:

For $k = 1, 2$,  $\textmd{ENC}_{{\sf sr},  {\sf it}, k}: \mathtt{C}^{2\times 1}\rightarrow \{0, 1\}$ maps ${\bf h}_k$ to $0$ or $1$ according to $\textmd{ENC}_{{\sf sr}, {\sf it}, k}\left({\bf h}_k\right)
	=
	{\bf 1}({\log_2\left(1 + P H_{k, k}\right) \geq 2\rho})$. Accordingly, receiver $k$ will send the feedback bit ``1''  if $\textmd{ENC}_{{\sf sr}, {\sf it}, k}\left({\bf h}_k\right) = 1$, and ``0'' otherwise. The decoder  $\textmd{DEC}_{{\sf sr}, {\sf it}}$ decodes the bits fed back by receivers and recovers the values of $\textmd{ENC}_{{\sf sr}, {\sf it}, k}\left({\bf h}_k\right)$ for $k = 1, 2$. The interference transmission pair $\left(p_1, p_2\right)$ is decided based on Table 1.

	\begin{table}[!htb]
	\caption{Decision rule of ${\textmd{\textit{DQ}}}_{{{{\sf sr}}}}$.}
\centering
\begin{tabular}{|c|c|c|}
  \hline
    $\textmd{ENC}_{{\sf sr}, {\sf it}, 1}\left({\bf h}_1\right)$ & $\textmd{ENC}_{{\sf sr}, {\sf it}, 2}\left({\bf h}_2\right)$ & $\left(p_1, p_2\right)$ \\
	\hline
  $1$ & $0$ & $\left(1, 0\right)$ \\
	\hline
  $0$ & $1$ & $\left(0, 1\right)$ \\
	\hline
  $1$ & $1$ & $\left(1, 0\right)$ or  $\left(0, 1\right)$\\
	\hline
  $0$ & $0$ & $\left(1, 1\right)$ \\
  \hline
\end{tabular}
\end{table}
Denote the network outage probability achieved by ${\textmd{\textit{DQ}}}_{{{{\sf sr}, {\sf it}}}}$ as $\textmd{OUT}\left({\textmd{\textit{DQ}}}_{{{{\sf sr}, {\sf it}}}}\right)$ and let $\textmd{FR}\left({\textmd{\textit{DQ}}}_{{{{\sf sr}, {\sf it}}}}\right)$ be the average feedback rate.\footnote{The average feedback rate in this paper is the sum of the average number of feedback bits fed back by each receiver.}

\begin{theorem}
$\textmd{\rm OUT}\left({\mathtt{\textit{DQ}}}_{{{{\sf sr}, {\sf it}}}}\right) = \textmd{\rm OUT}_{{\sf sr},  \sf it}^{{\sf opt}}$ and $\textmd{\rm FR}\left({\mathtt{\textit{DQ}}}_{{{{\sf sr}, {\sf it}}}}\right) = 2$.
\end{theorem}
\begin{IEEEproof}
With ${\textmd{\textit{DQ}}}_{{{{\sf sr}, {\sf it}}}}$, an outage event occurs only when $\textit{SR}_{\sf it}(p_1, p_2) < 2\rho$ for every $(p_1, p_2)\in\{\left(1, 0\right), \left(0, 1\right)$, $\left(1, 1\right)\},$ or equivalently when both receivers feeds back ``$0$'' and the corresponding power vector $\left(1, 1\right)$ from Table \Rmnum{1} still results in outage. This shows that $\textmd{\rm OUT}\left({\mathtt{\textit{DQ}}}_{{{{\sf sr}, {\sf it}}}}\right) = \textmd{\rm OUT}_{{\sf sr},  \sf it}^{{\sf opt}}$. Since two bits are fed back in total (one bit for either receiver), the average feedback rate is two bits per channel state.
	\end{IEEEproof}

The design of ${\textmd{\textit{DQ}}}_{{{{\sf sr}, {\sf it}}}}$ utilizes the fact that checking whether $(p_1, p_2) = \left(1, 0\right)$ or $\left(0, 1\right)$ leads to an outage event only requires the knowledge of local CSI at either receiver. Thus two bits of conferencing between receivers provides adequate information to each other for choosing the right pair $(p_1, p_2)$ to achieve the optimal performance.

We now consider the design of disributed quantizers for the time sharing strategy. In this case, we can similarly define the network outage probability of sum rate as $  \textmd{OUT}_{{\sf sr}, \sf ts} \triangleq \textmd{Pr}
\left\{
\textit{SR}_{\sf ts}\left(t_1, t_2\right) < 2\rho
\right\},\nonumber
$
where $\textit{SR}_{\sf ts}\left(t_1, t_2\right) \triangleq \sum_{k = 1}^2\textit{R}_{{\sf ts}, k} (t_k)$. Under the constraint of $t_1 + t_2$ = 1, the maximum sum rate can easily be calculated to be $\max\left\{\textit{SR}_{\sf ts}\left(1, 0\right), \textit{SR}_{\sf ts}\left(0, 1\right)\right\}$. Therefore, the optimal network outage probability is
\begin{align}
\textmd{OUT}_{{\sf sr}, \sf ts}^{ {\sf opt}} = \textmd{Pr}
\left\{
\textit{SR}_{\sf ts}\left(1, 0\right) < 2\rho, \textit{SR}_{\sf ts}\left(0, 1\right) < 2\rho
\right\}.\nonumber
\end{align}

Noticing that $\textit{SR}_{\sf ts}\left(1, 0\right) = \textit{SR}_{\sf it}\left(1, 0\right)$ and $ \textit{SR}_{\sf ts}\left(0, 1\right) = \textit{SR}_{\sf it}\left(0, 1\right) $ and using the same ideas as in the construction of $\textit{DQ}_{{\sf sr}, {\sf it}}$, we can design a  distributed quantizer for time sharing that achieves $\textmd{OUT}_{{\sf sr}, \sf ts}^{ {\sf opt}}$ with only one bit of feedback per receiver (we omit the details). On the other hand, the equalities $\textit{SR}_{\sf ts}(1, 0) = \textit{SR}_{\sf it}(1, 0)$ and $\textit{SR}_{\sf ts}(0, 1) = \textit{SR}_{\sf it}(0, 1)$ also imply $\textmd{OUT}_{{\sf sr}, \sf ts}^{ {\sf opt}} \leq \textmd{OUT}_{{\sf sr}, \sf it}^{ {\sf opt}}$. Hence, we only need to consider interference transmission if our objective is to minimize the network outage probability of the sum rate.

	\section{Distributed Quantization for Network Outage Probability of Minimum Rate}
\label{secfour}
We now study the design of distributed quantizers that minimize the outage probability of minimum rate. First, we determine the optimal network outage probability with time sharing or interference transmission. For time sharing, we define the network outage probability as
\begin{align}
  \textmd{OUT}_{{\sf mr}, {\sf ts}} \triangleq \textmd{Pr}\left\{\textit{MR}_{\sf ts}(t_1, t_2) < \rho\right \},\nonumber
\end{align}
where $\textit{MR}_{\sf ts}(t_1, t_2) \triangleq \min\left\{\textit{R}_{{\sf ts}, 1} (t_1), \textit{R}_{{\sf ts}, 2} (t_2)\right\}$ is the minimum achievable rate of the two transmitters. In interference transmission, the network outage probability is
\begin{align}
  \textmd{OUT}_{{\sf mr}, {\sf it}} \triangleq \textmd{Pr}\left\{\textit{MR}_{\sf it}(p_1, p_2) < \rho\right \},\nonumber
\end{align}
where $\textit{MR}_{\sf it}(p_1, p_2) \triangleq \min\left\{\textit{R}_{{\sf it}, 1} (p_1, p_2), \textit{R}_{{\sf it}, 2} (p_1, p_2)\right\}$. Now, let $(t_1^{\star},t_2^{\star}) = \arg\max_{(t_1,t_2)}\textit{MR}_{\sf ts}(t_1, t_2)$ and $(p_1^{\star},p_2^{\star}) = \arg\max_{(p_1,p_2)}\textit{MR}_{\sf it}(p_1, p_2)$ denote the optimal time sharing and power pairs that achieve $\textmd{OUT}_{{\sf mr}, {\sf ts}}$ and $\textmd{OUT}_{{\sf mr}, {\sf it}}$, respectively. We have the following two results, whose proofs can be found in Appendix A.
\begin{proposition}
We have
\begin{align}
\label{Optimal_Time_Sharing}
\begin{array}{l}
{t}_1^{\star}   =  \frac{ \log_2\left(1 + P H_{2, 2}\right)}{\log_2\left(1 + P H_{1, 1}\right) +\log_2\left(1 + P H_{2, 2}\right)},\\
{t}_2^{\star}  = \frac{ \log_2\left(1 + P H_{1, 1}\right)}{\log_2\left(1 + P H_{1, 1}\right) +\log_2\left(1 + P H_{2, 2}\right)}.
\end{array}
\end{align}
\end{proposition}
\begin{proposition}
If $\frac{P H_{1, 1}}{P H_{2, 1} + 1} \geq \frac{P H_{2, 2}}{P H_{1, 2} + 1}$, we have
	\begin{align}
	\label{First_p}
	({p}_1^{\star},p_2^{\star}) =  \textstyle \Biggl(\frac{\sqrt{\frac{4P^2 H_{1, 2} H_{2, 1} H_{2, 2} + 4P  H_{2, 2} H_{1, 2}}{H_{1, 1}} + 1} - 1}{2PH_{1, 2}}, 1\Biggr),
	\end{align}
	and otherwise, if $\frac{P H_{1, 1}}{P H_{2, 1} + 1} < \frac{P H_{2, 2}}{P H_{1, 2} + 1}$, we have
	\begin{align}
\label{Second_p}
	({p}_1^{\star},p_2^{\star}) = \textstyle \Biggl( 1, \frac{\sqrt{\frac{4 P^2 H_{1, 1} H_{1, 2} H_{2, 1}  + 4 P H_{1, 1}  H_{2, 1}}{H_{2, 2}} + 1} - 1}{2P H_{2, 1} }\Biggr).
	\end{align}
\end{proposition}

In particular, the optimal network outage probabilities of minimum rate for time sharing and interference transmission are given by $\textmd{OUT}_{{\sf mr}, \sf ts}^{ {\sf opt}}
 = \textmd{Pr}\left\{\textit{MR}_{\sf ts}(t_1^{\star}, t_2^{\star}) < \rho\right \}$ and $\textmd{OUT}_{{\sf mr}, \sf it}^{ {\sf opt}}
 = \textmd{Pr}\left\{\textit{MR}_{\sf it}(p_1^{\star}, p_2^{\star}) < \rho\right \}$, respectively.

We now propose two distributed quantizers, namely $\textmd{\textit{DQ}}_{{\sf mr}, {\sf ts}}$ and $\textmd{\textit{DQ}}_{{\sf mr}, {\sf it}}$. For the time sharing strategy, $\textmd{\textit{DQ}}_{{\sf mr}, {\sf ts}}$ will attain $\textmd{OUT}_{{\sf mr}, \sf ts}^{ {\sf opt}}$ exactly with a finite average feedback rate. For interference transmission, $\textmd{\textit{DQ}}_{{\sf mr}, {\sf it}}$ will approach $\textmd{OUT}_{{\sf mr}, \sf it}^{ {\sf opt}} $ tightly with a finite average feedback rate.

\subsection{Time Sharing}

 For a given ${\bf H}$, the minimum time percentage for receiver $k$ to prevent outage  is given by
  \begin{align}
    t_{k, \min} = \frac{\rho}{\log_2\left(1 + P H_{k, k}\right)},\nonumber
  \end{align}
  which can be calculated and known by receiver $k$, for $k = 1, 2$. Denote by $\textmd{\textit{DQ}}_{{\sf mr}, {\sf ts}}\left({\bf H}\right)$ the time sharing pair $(t_1, t_2)$ determined by $\textmd{\textit{DQ}}_{{\sf mr}, {\sf ts}}$.  The first task of $\textmd{\textit{DQ}}_{{\sf mr}, {\sf ts}}$ is to determine whether or not $\textit{MR}_{\sf ts}\left(t_1^{\star}, t_2^{\star}\right) \geq \rho$ through feedback communication between receivers. The first task is essentially a distributed decision-making problem. If $\textit{MR}_{\sf ts}\left(t_1^{\star}, t_2^{\star}\right) \geq \rho$ holds, the second task is to find $\textmd{\textit{DQ}}_{{\sf mr}, {\sf ts}}\left({\bf H}\right)$ that also enables $\textit{MR}_{\sf ts}\left(\textmd{\textit{DQ}}_{{\sf mr}, {\sf ts}}\left({\bf H}\right)\right) \geq \rho$.

The quantizer $\textmd{\textit{DQ}}_{{\sf mr}, {\sf ts}}$ is composed by two local encoders with the $k$th encoder $\textmd{ENC}_{{\sf mr}, {\sf ts}, k}$ located at receiver $k$ and a unique decoder  $\textmd{DEC}_{{\sf mr}, {\sf ts}}$ employed by all terminals.  We add the superscript ``$l$'' to indicate their operations in the $l$-th round of conferencing for $l \in \mathtt{N}$. Also, four parameters ${t}_{k, \min}^{\textmd{lb}}, {t}_{k, \min}^{\textmd{ub}}$ for $k = 1, 2$ are stored and updated at all terminals. Let ${t}_{k, \min}^{\textmd{lb}, l}, {t}_{k, \min}^{\textmd{ub}, l}$ represent the values of ${t}_{k, \min}^{\textmd{lb}}, {t}_{k, \min}^{\textmd{ub}}$ after round $l$.

 In round $0$, $\textmd{ENC}_{{\sf mr}, {\sf ts}, k}^{0}: \mathtt{C}^{2\times 1}\rightarrow \{0, 1\}$ maps ${\bf h}_k$ into $0$ or $1$ via $\textmd{ENC}_{{\sf mr}, {\sf ts}, k}^{0}\left({\bf h}_k\right)	=
	{\bf 1}(t_{k, \min} \geq 1)$, for $k = 1, 2$. Receiver $k$ will send the feedback bit ``1''  if $\textmd{ENC}_{{\sf mr}, {\sf ts}, k}^{0}\left({\bf h}_k\right) = 1$, and the feedback bit ``0'' otherwise. Then, $\textmd{DEC}_{{\sf mr}, {\sf ts}}^{0}$ decodes the bits fed back by receivers and recovers the values of $\textmd{ENC}_{{\sf mr}, {\sf ts}, k}^{0}\left({\bf h}_k\right) $ for $k = 1, 2$.  {If $\textmd{ENC}_{{\sf mr}, {\sf ts}, 1}^{0}\left({\bf h}_1\right)= 1$ or $\textmd{ENC}_{{\sf mr}, {\sf ts}, 2}^{0}\left({\bf h}_2\right) = 1$, an outage event is sure to happen. Then we set $\left(0.5, 0.5\right)$ as the time sharing pair (in fact, any time sharing pair can be used as outage is inavoidable) and the conferencing process ends.} Otherwise, ${t}_{k, \min}^{\textmd{lb}}$ and ${t}_{k, \min}^{\textmd{ub}} = 1$ are updated as ${t}_{k, \min}^{\textmd{lb}, 0} = 0, {t}_{k, \min}^{\textmd{ub}, 0} = 1$ for $k = 1, 2$, then $\textmd{\textit{DQ}}_{{\sf mr}, {\sf ts}}$ continues to the next round.

In round $l$ where $l \in \mathtt{N} - \{0\}$, $\textmd{ENC}_{{\sf mr}, {\sf ts}, k}^{l}: \mathtt{C}^{2\times 1}\rightarrow \{0, 1\}$ maps ${\bf h}_k$ into $0$ or $1$ according to
\begin{align}
\textmd{ENC}_{{\sf mr}, {\sf ts}, k}^{l}\left({\bf h}_k\right)
	=
	{\bf 1}\left(t_{k, \min} \geq \textstyle \frac{{t}_{k, \min}^{\textmd{lb}, l - 1} + {t}_{k, \min}^{\textmd{ub}, l - 1}}{2}\right),\nonumber
\end{align}
for $k = 1, 2$. Receiver $k$ will send 1 bit of ``1''  if $\textmd{ENC}_{{\sf mr}, {\sf ts}, k}^{l}\left({\bf h}_k\right) = 1$, and ``0'' otherwise. Then $\textmd{DEC}_{{\sf mr}, {\sf ts}}^{l}$ decodes the bits fed back by receivers and recovers the values of $\textmd{ENC}_{{\sf mr}, {\sf ts}, k}^{l}\left({\bf h}_k\right) $ for $k = 1, 2$.
\begin{enumerate}
	\item If $\textmd{ENC}_{{\sf mr}, {\sf ts}, 1}^{l}\left({\bf h}_1\right) = \textmd{ENC}_{{\sf mr}, {\sf ts}, 2}^{l}\left({\bf h}_2\right) = 1$, an outage event is inavoidable. We thus set $(0.5, 0.5)$ as the time sharing pair and conferencing ends.
	\item If $\textmd{ENC}_{{\sf mr}, {\sf ts}, 1}^{l}\left({\bf h}_1\right) = \textmd{ENC}_{{\sf mr}, {\sf ts}, 2}^{l}\left({\bf h}_2\right) = 0$, we set $\textmd{\textit{DQ}}_{{\sf mr}, {\sf ts}}\left({\bf H}\right) = \left(\frac{{t}_{1, \min}^{\textmd{lb}, l -1} + {t}_{1, \min}^{\textmd{ub}, l - 1}}{2}, \frac{{t}_{2, \min}^{\textmd{lb}, l - 1} + {t}_{2, \min}^{\textmd{ub}, l - 1}}{2}\right)$ as the time sharing pair, and conferencing ends.
	\item If $\textmd{ENC}_{{\sf mr}, {\sf ts}, 1}^{l}\left({\bf h}_1\right) =1$ and $ \textmd{ENC}_{{\sf mr}, {\sf ts}, 2}^{l}\left({\bf h}_2\right) = 0$,  we let $t_{1, \min}^{\textmd{ lb}, l} = \frac{{t}_{1, \min}^{\textmd{lb}, l - 1} + {t}_{1, \min}^{\textmd{ub}, l - 1}}{2}$ and $t_{2, \min}^{\textmd{ub}, l} = \frac{{t}_{2, \min}^{\textmd{lb}, l - 1} + {t}_{2, \min}^{\textmd{ub}, l - 1}}{2}$. If $\textmd{ENC}_{{\sf mr}, {\sf ts}, 1}^{l}\left({\bf h}_1\right) =0$ and  $\textmd{ENC}_{{\sf mr}, {\sf ts}, 2}^{l}\left({\bf h}_2\right) = 1$, we let $t_{1, \min}^{\textmd{ ub}, l} = \frac{{t}_{1, \min}^{\textmd{lb}, l - 1} + {t}_{1, \min}^{\textmd{ub}, l - 1}}{2}$ and $t_{2, \min}^{\textmd{lb}, l} = \frac{{t}_{2, \min}^{\textmd{lb}, l - 1} + {t}_{2, \min}^{\textmd{ub}, l - 1}}{2}$. In either case, conferencing continues to the next round.
\end{enumerate}

 Note that the condition $\textmd{MR}_{\sf ts}\left(\textmd{\textit{DQ}}_{{\sf mr}, {\sf ts}}\left({\bf H}\right)\right) < \rho$ is equivalent to $t_{1, \min} + t_{2,  \min} > 1$, and $\textmd{\textit{DQ}}_{{\sf mr}, {\sf ts}}$ determines whether $t_{1, \min} + t_{2,  \min} > 1$ holds or not. To accomplish this, either receiver quantizes its own $t_k$ in a finer and finer way when $l$ increases and tells the quantized feedback bits to others. The parameters ${t}_{k, \min}^{\textmd{lb}}, {t}_{k, \min}^{\textmd{ub}}$ serve as the lower and upper bounds on $t_{k, \min}$ updated by conferencing between receivers. The decision of whether $t_{1, \min} + t_{2,  \min} > 1$ holds or not is made by jointly considering ${t}_{k, \min}^{\textmd{lb}}$ and $ {t}_{k, \min}^{\textmd{ub}}$. The inter-receiver conferencing process will continue until the exchanged feedback bits are adequate to make a precise decision about whether $t_{1, \min} + t_{2,  \min} > 1$ holds or not.

Let $\textmd{OUT}\left(\textmd{\textit{DQ}}_{{\sf mr}, {\sf ts}}\right)$ and $\textmd{FR}\left(\textmd{\textit{DQ}}_{{\sf mr}, {\sf ts}}\right)$ denote the network outage probability and  average feedback rate of $\textmd{\textit{DQ}}_{{\sf mr}, {\sf ts}}$, respectively. The following theorem shows that whenever the optimal time shairing pair $(t_1^{\star}, t_2^{\star})$ in Proposition 1 can avoid outage, the time sharing pair picked by $\textmd{\textit{DQ}}_{{\sf mr}, {\sf ts}}$ will also avoid outage with probability one, and that the average feedback rate of $\textmd{\textit{DQ}}_{{\sf mr}, {\sf ts}}$ is finite. The proof is provided in Appendix B.

\begin{theorem}
For any $P > 0$, we have
\begin{align}
\textmd{\rm OUT}\left(\mathtt{\textit{DQ}}_{{\sf mr}, {\sf ts}}\right) = \textmd{\rm OUT}_{{\sf mr}, \sf ts}^{ {\sf opt}},
\end{align}
and
\begin{align}
\label{FR_Time_Sharing}
\textmd{\rm FR}\left(\mathtt{\textit{DQ}}_{{\sf mr}, {\sf ts}}\right) \leq 2 + 2 e^{-\frac{\rho \log 2}{P}}\left(1 + \frac{C_0}{P}\right),
\end{align}
where $C_0$ is a bounded constant that is independent of $P$.\footnote{Since we focus on showing the average feedback rate is finite for any $P$, it is beyond the scope of our paper to derive the tightest bound, i.e., the smallest value for $C_0$. }
\end{theorem}

Theorem 2 shows zero-distortion in network outage probability actually can be achieved by finite average feedback rates, other than infinite number of feedback bits in the traditional view. This surprising result comes from our design for feedback communication between receivers based on conferencing.
%
%\textit{Remark 2:} We can also find from \eqref{FR_Time_Sharing} that when $P\rightarrow \infty$ or $P\rightarrow 0$, the average feedback rate will be nearly 4 or 2, respectively. This can be intuitively interpreted as follows: when $P\rightarrow \infty$, the probability that $t_{k, \min} < \frac{1}{2}$ for $k = 1, 2$ is increasing towards one.  $\textmd{\textit{DQ}}_{{\sf mr}, {\sf ts}}$ will quit after 2 rounds with $\textmd{\textit{DQ}}_{{\sf mr}, {\sf ts}}\left({\bf H}\right) = (0.5, 0.5)$, thus the average feedback rate is $4$; when
%$P\rightarrow 0$, the probability that $t_{k, \min}> 1$ for $k = 1, 2$ is also almost one, thus after round 0, $\textmd{\textit{DQ}}_{{\sf mr}, {\sf ts}}$ will quit and the average feedback rate is 2.

\subsection{Interference Transmission}

For $k, l = 1, 2$ and $k \neq l$,  the maximum allowed power of transmitter $k$ that will not cause outage to receiver $l$ when transmitter $l$ uses full power can be calculated to be
\begin{align}
  p_{k, \max} = \frac{H_{l, l}}{\left(2^{{\rho}}-1\right) H_{k, l}} - \frac{1}{P H_{k, l}}.\nonumber
\end{align}
Note that $p_{k, \max}$ can be calculated at receiver $l$.

The proposed quantizer $\textmd{\textit{DQ}}_{{\sf mr}, {\sf it}}$ consists of two local encoders, two local compressors and a unique decoder. The $k$-th encoder $\textmd{ENC}_{{\sf mr}, {\sf it}, k}$ and $k$-th compressor $\textmd{CMP}_{{\sf mr}, {\sf it}, k}$ are located at receiver $k$, while the decoder  $\textmd{DEC}_{{\sf mr}, {\sf it}}$ is used by all terminals. We add the superscript ``$l$'' to indicate their operations in the $l$-th round of conferencing for $l = 0, 1$.

For any $M \in \mathtt{N} - \{0\}$, let $\mathcal{C}_M = \left\{\frac{m}{M}: m = 0, \ldots, M\right\}$. Denote $\textmd{\textit{DQ}}_{{\sf mr}, {\sf it}}\left({\bf H}\right)$ as the interference transmission pair $(p_1, p_2)$ determined by $\textmd{\textit{DQ}}_{{\sf mr}, {\sf it}}$. There are at most two rounds of conferencing in $\textmd{\textit{DQ}}_{{\sf mr}, {\sf it}}$.

 In round $0$, $\textmd{ENC}_{{\sf mr}, {\sf it}, 1}^{0}: \mathtt{C}^{2\times 1}\rightarrow \mathcal{C}_M$ maps ${\bf h}_1$ into a codeword in $\mathcal{C}_M$ according to
      \begin{align}
      \textmd{ENC}_{{\sf mr}, {\sf it}, 1}^{0} \left({\bf h}_1\right) = \left\{
      \begin{matrix}
      0, &{p}_{2, \max} \leq 0,\\
      \argmax\limits_{x \in \mathcal{C}_M, x \leq {p}_{2, \max} } x, & {p}_{2, \max} >0.
      \end{matrix}
      \right.
      \nonumber
      \end{align}
      Then $\textmd{CMP}_{{\sf mr}, {\sf it}, 1}^{0}: \mathcal{C}_M \rightarrow \mathcal{B}$ maps the index of $\textmd{ENC}_{{\sf mr}, {\sf it}, 1}^{0} \left({\bf h}_1\right)$ to a binary description in $\mathcal{B}$, a set of binary representations for codewords in $\mathcal{C}$. With fixed-length coding, $\left\lceil \log_2\left|\mathcal{C}\right|\right\rceil = \left\lceil \log_2(M + 1)\right\rceil$ bits indicating the index of $\textmd{ENC}_{{\sf mr}, {\sf it}, 1}^{0} \left({\bf h}_1\right)$ are fed back by receiver 1.\footnote{The performance of $\textmd{\textit{DQ}}_{{\sf mr}, {\sf it}}$ can be improved by taking variable-length coding into consideration. We use fixed-length coding here for convenience.} $\textmd{DEC}_{{\sf mr}, {\sf it}}^{0}$ decodes them and recovers the value of $\textmd{ENC}_{{\sf mr}, {\sf it}, 1}^{0} \left({\bf h}_1\right)$, then receiver 2 will send one bit of ``1'' if $\log_2\left(1 + \frac{ \textmd{ENC}_{{\sf mr}, {\sf it}, 1}^{0} \left({\bf h}_1\right) P H_{2, 2}}{ P H_{1, 2} + 1}\right) \geq \rho$, and ``0'' otherwise. If ``1'' is fed back by receiver 2, $\textmd{\textit{DQ}}_{{\sf mr}, {\sf it}}\left({\bf H}\right) = \left(1, \textmd{ENC}_{{\sf mr}, {\sf it}, 1}^{0} \left({\bf h}_1\right) \right)$ is the decided pair and thus, conferencing for the current channel state finishes. Otherwise, conferencing will continue to the next round.

      In round $1$, $\textmd{ENC}_{{\sf mr}, {\sf it}, 2}^{1}: \mathtt{C}^{2\times 1}\rightarrow \mathcal{C}_M$ maps ${\bf h}_2$ into a codeword in $\mathcal{C}_M$ according to
      \begin{align}
      \textmd{ENC}_{{\sf mr}, {\sf it}, 2}^{1} \left({\bf h}_2\right) = \left\{
      \begin{matrix}
      0, &{p}_{1, \max} \leq 0,\\
      \argmax\limits_{x \in \mathcal{C}_M, x \leq {p}_{1, \max} } x, & {p}_{1, \max} >0.
      \end{matrix}
      \right.
      \nonumber
      \end{align}
			 Then $\textmd{CMP}_{{\sf mr}, {\sf it}, 2}^{1}: \mathcal{C}_M \rightarrow \mathcal{B}$ maps the index of $\textmd{ENC}_{{\sf mr}, {\sf it}, 2}^{1} \left({\bf h}_2\right)$ to a binary description in $\mathcal{B}$. $\left\lceil \log_2(M + 1)\right\rceil$ bits indicating the index of $\textmd{ENC}_{{\sf mr}, {\sf it}, 2}^{1} \left({\bf h}_2\right)$ are fed back by receiver 2. $\textmd{DEC}_{{\sf mr}, {\sf it}}^{1}$ decodes them and recovers the value of $\textmd{ENC}_{{\sf mr}, {\sf it}, 2}^{1} \left({\bf h}_2\right)$, and $\textmd{\textit{DQ}}_{{\sf mr}, {\sf it}}\left({\bf H}\right) = \left(\textmd{ENC}_{{\sf mr}, {\sf it}, 2}^{1} \left({\bf h}_2\right), 1 \right)$ is the final interference transmission pair.
			
			 The interference transmission pair decided by $\textmd{\textit{DQ}}_{{\sf mr}, {\sf it}}$ has at least one element equal to $1$, i.e., $p_1 = 1$ or $p_2 = 1$, which arises from the fact that the performance of any pair that does not satisfy this can be improved by multiplying the pair with a scaling factor until at least one element reaches $1$  \cite{OptimalPowerControlSumRate}. Therefore, the proposed quantizer only needs to work on the non-one element. To do this, either receiver tries to tell others the maximum power it can tolerate for preventing outage.
			
			 Denote the network outage probability and average feedback rate of $\textmd{\textit{DQ}}_{{\sf mr}, {\sf it}}$ by $\textmd{OUT}\left(\textmd{\textit{DQ}}_{{\sf mr}, {\sf it}}\right)$ and $\textmd{FR}\left(\textmd{\textit{DQ}}_{{\sf mr}, {\sf it}}\right)$, respectively. The following theorem provides upper bounds on $\textmd{OUT}\left(\textmd{\textit{DQ}}_{{\sf mr}, {\sf it}}\right)$ and $\textmd{FR}\left(\textmd{\textit{DQ}}_{{\sf mr}, {\sf it}}\right)$. The proof of the theorem is provided in Appendix D.
			
\begin{theorem}
For any $P>0$ and $M \in \mathtt{N} - \{0\}$, we have
\begin{align}
\label{DQ_OUT_OUT}
\textmd{\rm OUT}\left(\textmd{\textit{DQ}}_{{\sf mr}, {\sf it}}\right)
\leq \textmd{\rm OUT}_{{\sf mr}, \sf it}^{ {\sf opt}}  + \frac{C_1}{M},
\end{align}
and
\begin{align}
\label{AFR_NOP}
\textmd{\rm FR}\left(\textmd{\textit{DQ}}_{{\sf mr}, {\sf it}}\right) \leq 2\log_2(M + 1) + 3,
\end{align}
where $C_1 > 0$ is a bounded constant that is independent of $P$ and $M$.
\end{theorem}

 From Theorem 3, it is seen that the distortion in network outage probability is inversely proportional to $M$, while the average feedback rate is bounded by a finite constant plus the term $2\log_2(M + 1)$ that scales as $O\left(\log(M)\right)$. Letting $M$ satisfy $2\log_2(M + 1) + 3 = \textit{R}$, we can observe that the loss in outage probability due to quantization decays at least exponentially with the total feedback rate $\textit{R}$ as $O\left(2^{-\frac{\textit{R}}{2}}\right)$.

\subsection{Time Sharing or Interference Transmission?}

We recall from Section \Rmnum{3} that for the network outage probability of sum rate, the interference transmission is always superior to time sharing. On the other hand, for the network outage probability of minimum rate, depending on the power constraing $P$, either one of two transmission strategies may be optimal. To illustrate this phenomenon, the network outage probabilities $\textmd{OUT}_{{\sf mr}, \sf ts}^{ {\sf opt}} $ and $\textmd{OUT}_{{\sf mr}, \sf it}^{ {\sf opt}}$ are plotted  versus $P$ for various $\epsilon$ in Fig. 1. The target data rate is $\rho = 0.5$. We can observe from Fig. 1 that for any given $\epsilon$, there is a threshold power level $P_{\textmd{th}}$ (that depends on $\epsilon$) such that when $P \leq P_{\textmd{th}}$, $\textmd{OUT}_{{\sf mr}, \sf ts}^{ {\sf opt}}  \leq \textmd{OUT}_{{\sf mr}, \sf it}^{ {\sf opt}}$, and when $P > P_{\textmd{th}}$, $\textmd{OUT}_{{\sf mr}, \sf ts}^{ {\sf opt}}  > \textmd{OUT}_{{\sf mr}, \sf it}^{ {\sf opt}}$. In other words, we should use interference transmission when $P \leq P_{\textmd{th}}$, and otherwise, if $P > P_{\textmd{th}}$, we should utilize the time sharing strategy. The decision between time sharing and interference transmission only requires the knowledge of $P_{\textmd{th}}$, which can be a prior information known by all terminals. Although it is difficult to derive a closed-form expression of $P_{\textmd{th}}$, it can still be estimated through numerical simulations. For example, according to Fig. 1, we have $P_{\textmd{th}}\approx 2, 5, 12, 25$ dB  when $\epsilon = 1, 0.5, 0.1$ and $0.01$, respectively.

\begin {figure}
\centering
\includegraphics[width= 4.5 in]{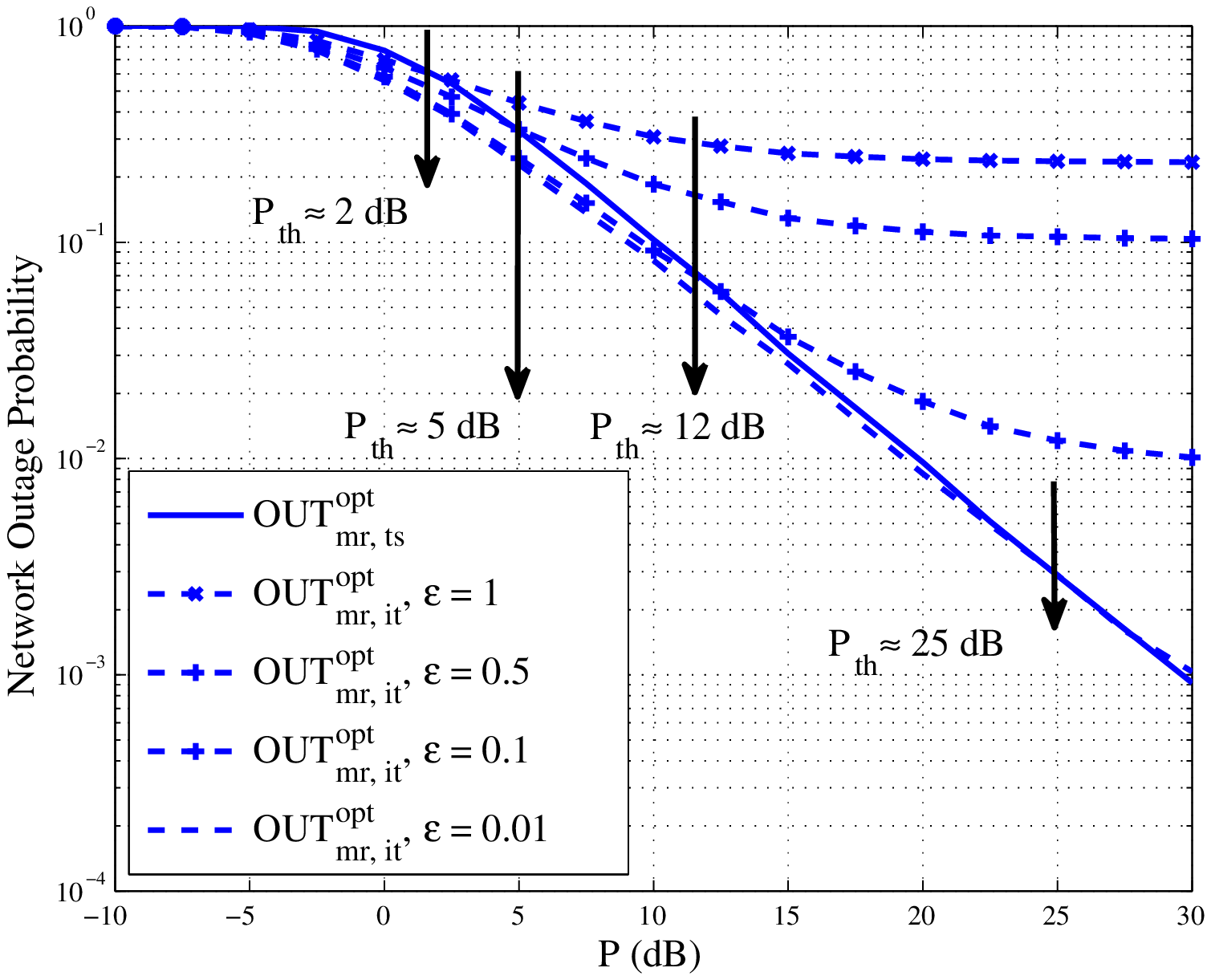}
\caption{$\textmd{OUT}_{{\sf mr}, \sf ts}^{ {\sf opt}} $ and $\textmd{OUT}_{{\sf mr}, \sf it}^{ {\sf opt}}$ versus $P$.}
\label{fig1, 2}
\end{figure}

\section{Numerical Simulations}
\label{secnumresults}

\begin {figure}
\centering
\includegraphics[width= 4.5 in]{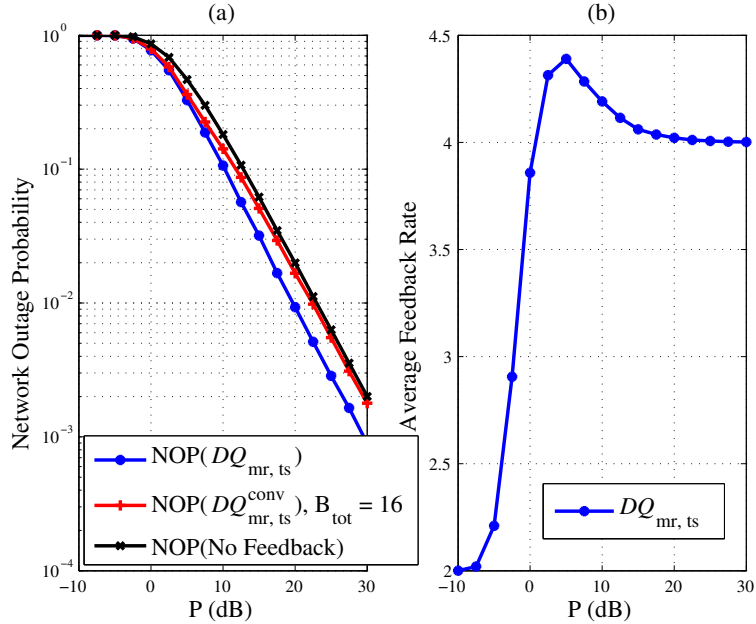}
\caption{Simulated network outage probabilities of minimum rate for $\textit{DQ}_{{\sf mr}, {\sf ts}}$, $\textit{DQ}_{{\sf mr}, {\sf ts}}^{\sf conv}$ and the case with no feedback as well as the average feedback rate of $\textit{DQ}_{{\sf mr}, {\sf ts}}$ versus $P$.}
\end{figure}

In this section, we present simulations to verify the theoretical results for $\textit{DQ}_{{\sf mr}, {\sf ts}}$ in time sharing and $\textit{DQ}_{{\sf mr}, {\sf it}}$ in interference transmission. For each instance of $P$ and $\epsilon$, a sufficient number of channel state realizations are generated to observe at least 5000 outage events. We have chosen $\rho = 0.5$.

We will compare the performance of the proposed quantizers with that of the conventional one \cite{Interference_Power_Control, Interference_Throughput} denoted by $\textit{DQ}_{{\sf mr}}^{\sf conv}$ in time sharing and interference transmission, respectively. For readers' convenience, we provide a brief description of the quantizer $\textit{DQ}_{{\sf mr}}^{\sf conv}$ as described in \cite{Interference_Power_Control, Interference_Throughput}. For $k = 1, 2$, receiver $k$ employs $\frac{B_{\sf tot}}{4}$ bits to quantize $H_{1, k}$ and $H_{2, k}$  separately based on a scalar codebook generated by  Lloyd Algorithm \cite{GLA} with the cardinality being $2^{\frac{B_{\sf tot}}{4}}$. All terminals decode the feedback bits and reconstruct the quantized $\bf H$ as $\hat{\bf H}$. In time sharing, ${t}_1^{\star}$ and ${t}_2^{\star}$ are calculated according to Proposition 1 by treating $\hat{\bf H}$ as ${\bf H}$, while in interference transmission, ${p}_1^{\star}$ and ${p}_2^{\star}$ are computed by Proposition 2 based on $\hat{\bf H}$. The average feedback rate of $\textit{DQ}_{{\sf mr}}^{\sf conv}$ is $B_{\sf tot}$ bits per channel state. We add the subscript of ``$\sf ts$'' or ``$\sf it$'' to $\textit{DQ}_{{\sf mr}}^{\sf conv}$ to distinguish when it is applied in time sharing or interference transmission, respectively.

In Fig. 2 (a), the network outage probabilities of minimum rate for $\textit{DQ}_{{\sf mr}, {\sf ts}}$, $\textit{DQ}_{{\sf mr}, {\sf ts}}^{\sf conv}$ (with $B_{\sf tot} = 16$) and the case with no feedback (where either transmitter consumes half of the entire block to transmit, i.e., $t_1 = t_2 = 0.5$) are plotted. It is shown that the network outage probabilities of the latter two scenarios are worse than that of $\textit{DQ}_{{\sf mr}, {\sf ts}}$ (the minimum one), which substantiates that feedback is necessary as well as the proposed quantizer based on conferencing is superior.  Fig. 2 (b) plots the average feedback rate of $\textit{DQ}_{{\sf mr}, {\sf ts}}$, which is finite and small in the entire interval of $P$.  Furthermore, when $P\rightarrow \infty$ or $0$, the average feedback rate approaches towards $4$ or $2$, respectively. This corresponds to the upper bound in Theorem 2 and it can be intuitively interpreted like this: when $P\rightarrow \infty$, the probability that $t_{k, \min} < \frac{1}{2}$ for $k = 1, 2$, is increasing towards $1$, then after two rounds,  $\left(0.5, 0.5\right)$ will be chosen as $\textit{DQ}_{{\sf mr}, {\sf ts}}\left({\bf H}\right)$ most likely. On the other hand, when $P\rightarrow 0$, the probability that $t_{k, \min} > 1$ for $k = 1, 2$, also goes to $1$, thus after round $0$, the quantization process will finish because an outage event is inevitable almost surely.

\begin {figure}
\centering
\includegraphics[width= 4.5 in]{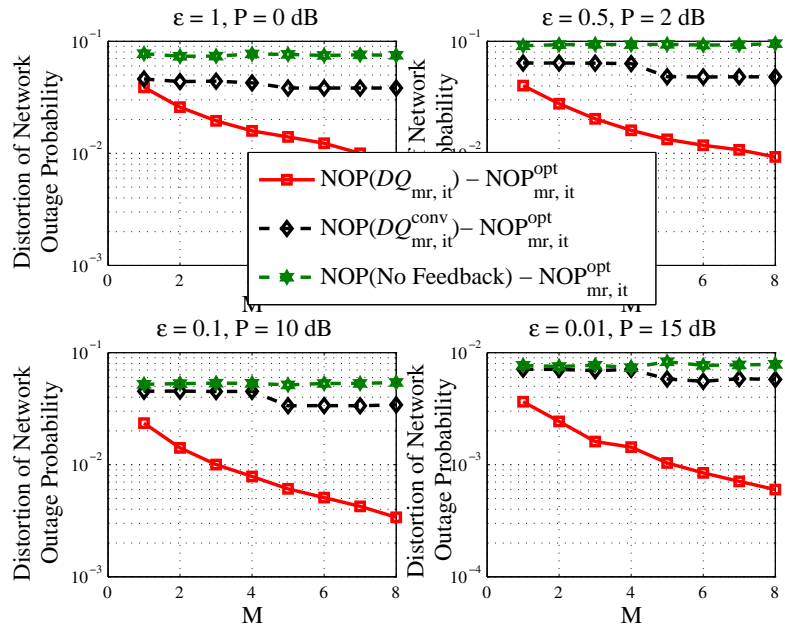}
\caption{Distortions of network outage probability for minimum rate of $\textit{DQ}_{{\sf mr}, {\sf it}} $, $\textit{DQ}_{{\sf mr}, {\sf it}}^{\sf conv}$ and the case with no feedback versus $M$.}
\end{figure}

In Fig. 3, we show the distortions of network outage probability for minimum rate of $\textit{DQ}_{{\sf mr}, {\sf it}} $, $\textit{DQ}_{{\sf mr}, {\sf it}}^{\sf conv}$ and the case with no feedback (where both transmitters will use full power, i.e., $p_1 = p_2 = 1$) versus $M$. For each $\epsilon$, we choose a value of $P$ smaller than $P_{\textmd{th}}$ thus interference transmission should be applied.  In order to demonstrate that $\textit{DQ}_{{\sf mr}, {\sf it}}$ outperforms $\textit{DQ}_{{\sf mr}, {\sf it}}^{\sf conv}$ even when $\textit{DQ}_{{\sf mr}, {\sf it}}^{\sf conv}$ has a higher feedback rate, we choose the number of feedback bits assigned to $\textit{DQ}_{{\sf mr}, {\sf it}}^{\sf conv}$ as is $B_{\sf tot} = 4 \left\lceil \frac{2\log_2 (M + 1) + 3}{4} \right\rceil $. Note that $B_{\sf tot} = 8$ when $1 \leq M \leq 4$ and $12$ when $5\leq M \leq 8$. The distortions of $\textit{DQ}_{{\sf mr}, {\sf it}}$ and $\textit{DQ}_{{\sf mr}, {\sf it}}^{\sf conv}$ versus both $P$ and the average feedback rate are also shown in Fig. 4 for different values of $\epsilon$. It can be observed that in  interference transmission, (i) the distortion of $\textit{DQ}_{{\sf mr}, {\sf it}}$ decreases almost linearly with increasing $M$ in the $\log$-scale, which corresponds to the upper bound derived in Theorem 3; (ii) the decreasing speed of the distortion for $\textit{DQ}_{{\sf mr}, {\sf it}}$ in regard to $M$ or the average feedback rate is much faster than that of $\textit{DQ}_{{\sf mr}, {\sf it}}^{\sf conv}$;  (ii) the distortion of $\textit{DQ}_{{\sf mr}, {\sf it}}$ is much smaller than those of $\textit{DQ}_{{\sf mr}, {\sf it}}^{\sf conv}$ and the case with no feedback, which verifies that feedback is necessary and our proposed distributed quantizer based on conferencing outperforms the conventional distributed quantizer.

\begin {figure}
\centering
\includegraphics[width= 4.5 in]{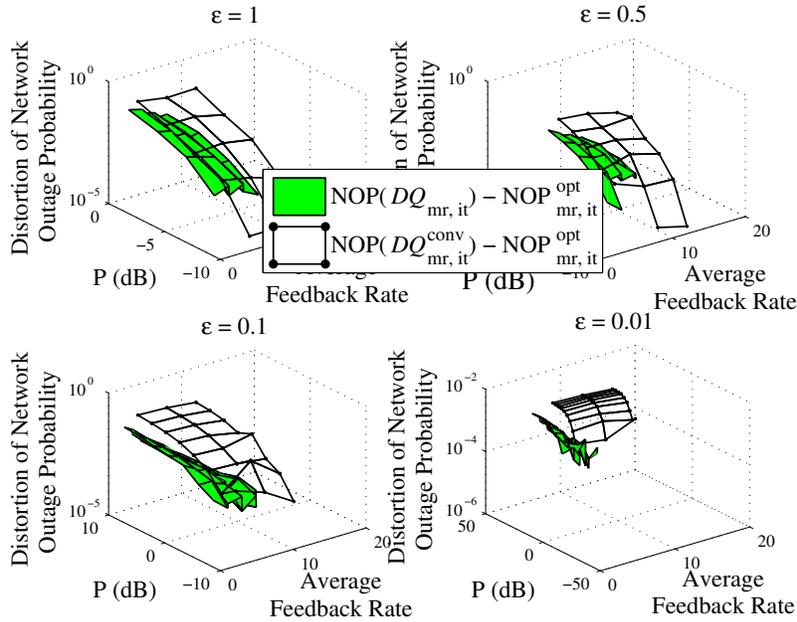}
\caption{Distortions of network outage probability for minimum rate of $\textit{DQ}_{{\sf mr}, {\sf it}} $ and $\textit{DQ}_{{\sf mr}, {\sf it}}^{\sf conv}$ versus $P$ and average feedback rate.}
\end{figure}

\section{Conclusions and Future Work}

We have introduced conferencing-based distributed channel quantizers for a two-user interference network where interference signals are treated as noise. We have shown that the proposed distributed quantizers can achieve or closely approach the optimal network outage probabilities of sum rate and minimum rate in time sharing or interference transmission with finite average feedback rates.

So far, we have studied the scenario where only one transmission strategy (interference transmission or time sharing) is used for every channel state. We note that utilizing different transmission strategies for different channel states will result in a better performance. The design and analysis of distributed quantizers for such an adaptive system is an interesting future research direction.

\section*{Acknowledgement}
This work was supported in part by the NSF Award CCF-1218771.

\appendices
\section{Proofs of Propositions 1 and 2}
\begin{IEEEproof}
The optimal time sharing pair $(t_1^{\star}, t_2^{\star})$ that minimizes $\textmd{OUT}_{\sf ts}^{\sf mr}$ also maximizes $\textit{MR}_{\sf ts}(t_1, t_2)$. Substituting $t_2 = 1 - t_1$ into $\textit{MR}_{\sf ts}(t_1, t_2)$, the problem that maximizes $\textit{MR}_{\sf ts}(t_1, t_2)$ becomes $\maxmin\limits_{0 \leq t_1\leq 1}\left\{  t_1\log_2\left(1 +  P {H_{1, 1}}\right), (1 - t_1)\log_2\left(1 +  P {H_{2, 2}}\right) \right\}$.  The first term is increasing in $t_1$ while the second term is decreasing in $t_1$. Therefore, the maximum is reached when $t_1\log_2\left(1 +  P {H_{1, 1}}\right) = (1 - t_1)\log_2\left(1 +  P {H_{2, 2}}\right)$, yielding $t_1^{\star}$ and $t_2^{\star}$ given in \eqref{Optimal_Time_Sharing}.

The optimal interference transmission pair $(p_1^{\star}, p_2^{\star})$ that minimizes $\textmd{OUT}_{\sf it}^{\sf mr}$ also maximizes $\textit{MR}_{\sf it}(p_1, p_2)$. We first show $p_1^{\star} = 1$ or $p_2^{\star} = 1$. Assume by contradiction that $0< p_1^{\star}, p_2^{\star} < 1$. Let $\beta = \min\left\{\frac{1}{p_1^{\star}}, \frac{1}{p_2^{\star}}\right\} > 1$, then
\begin{align}
\label{Optimality_One}
{{\textit{MR}}}_{\sf it}\left(\beta p_1^{\star}, \beta p_2^{\star}\right)
& =\min
\left\{
\log_2\left(1 + \frac{P \beta p_1^{\star} H_{1, 1}}{P \beta p_2^{\star} H_{2, 1} + 1}\right),
\log_2\left(1 + \frac{P \beta p_2^{\star} H_{2, 2}}{P \beta p_1^{\star} H_{1, 2} + 1}\right)
\right\}
\nonumber\\
&=\min
\left\{
\log_2\left(1 + \frac{P  p_1^{\star} H_{1, 1}}{P p_2^{\star} H_{2, 1} + \frac{1}{\beta}}\right),
\log_2\left(1 + \frac{P p_2^{\star} H_{2, 2}}{P p_1^{\star} H_{1, 2} + \frac{1}{\beta}}\right)
\right\}
\nonumber\\
&>
\min
\left\{
\log_2\left(1 + \frac{P  p_1^{\star} H_{1, 1}}{P p_2^{\star} H_{2, 1} + 1}\right),
\log_2\left(1 + \frac{P p_2^{\star} H_{2, 2}}{P p_1^{\star} H_{1, 2} + 1}\right)
\right\}
\nonumber\\
& ={{\textit{MR}}}_{\sf it}\left(p_1^{\star}, p_2^{\star}\right),
\end{align}
which contradicts the assumption that $\left(p_1^{\star}, p_2^{\star}\right)$ is optimal. Therefore, $p_1^{\star} = 1$ or $p_2^{\star} = 1$.

When $p_1^{\star} = 1$, the problem that maximizes $\textit{MR}_{\sf it}\left(p_1, p_2\right)$ is equivalent to $\maxmin\limits_{0< p_2 \leq 1}
\left\{
\frac{P  H_{1, 1}}{P  p_2 H_{2, 1} + 1},
\frac{P  p_2 H_{2, 2}}{P  H_{1, 2} + 1}
\right\}$, where $\frac{P  H_{1, 1}}{P  p_2 H_{2, 1} + 1}$ is decreasing in $p_2$ and $\frac{P  p_2 H_{2, 2}}{P  H_{1, 2} + 1}$ is increasing in $p_2$. Letting $\frac{P  H_{1, 1}}{P  p_2 H_{2, 1} + 1} = \frac{P p_2 H_{2, 2}}{P H_{1, 2} + 1}$, the positive root is  $\tilde{p}_2 = \frac{\sqrt{\frac{4 P^2 H_{1, 1} H_{1, 2} H_{2, 1}  + 4 P H_{1, 1}  H_{2, 1}}{H_{2, 2}} + 1} - 1}{2P H_{2, 1} }$. Note that $0<\tilde{p}_2  < 1$ holds only when $\frac{P  H_{1, 1}}{P H_{2, 1} + 1} <\frac{H_{2, 2}}{P H_{1, 2} + 1}$. Thus, when $\frac{P  H_{1, 1}}{P H_{2, 1} + 1} <\frac{H_{2, 2}}{P H_{1, 2} + 1}$, $p_1^{\star} = 1$ and $p_2^{\star} = \tilde{p}_2 $.  Similarly, when $p_2^{\star} = 1$,  we derive the positive root of $\frac{P p_1 H_{1, 1}}{P  H_{2, 1} + 1} = \frac{P  H_{2, 2}}{P p_1 H_{1, 2} + 1}$ as $\tilde{p}_1 = \frac{\sqrt{\frac{4P^2 H_{1, 2} H_{2, 1} H_{2, 2} + 4P  H_{2, 2} H_{1, 2}}{H_{1, 1}} + 1} - 1}{2PH_{1, 2}}$. Note that $0< \tilde{p}_1 < 1$ holds when $\frac{P  H_{1, 1}}{P H_{2, 1} + 1} \geq \frac{H_{2, 2}}{P H_{1, 2} + 1}$. Hence, when $\frac{P  H_{1, 1}}{P H_{2, 1} + 1} \geq \frac{H_{2, 2}}{P H_{1, 2} + 1}$, $p_1^{\star} = \tilde{p}_1$ and $p_2^{\star} = 1$.
\end{IEEEproof}

\section{Proof of Theorem 2}
\begin{IEEEproof}
Let
\begin{align}
\begin{array}{l}
 \mathcal{H}_{1}   = \left\{{\bf H}: t_{1, \min} + t_{2, \min} > 1, t_{1, \min}, t_{2, \min} > 0\right\},\nonumber\\
  \mathcal{H}_{2}   = \left\{{\bf H}: t_{1, \min} + t_{2, \min} = 1, t_{1, \min}, t_{2, \min} > 0\right\}, \nonumber\\
   \mathcal{H}_{3}   = \left\{{\bf H}: t_{1, \min} + t_{2, \min} < 1, t_{1, \min}, t_{2, \min} > 0\right\}.\nonumber
\end{array}
\end{align}
Note that $t_{1, \min} + t_{2, \min} = \frac{\rho}{\log_2\left(1 + P H_{1, 1}\right)} + \frac{\rho}{\log_2\left(1 + P H_{2, 2}\right)} = \frac{\rho}{\frac{{\log_2\left(1 + P H_{1, 1}\right)} {\log_2\left(1 + P H_{2, 2}\right)}}{{\log_2\left(1 + P H_{1, 1}\right)} + {\log_2\left(1 + P H_{2, 2}\right)}}} = \frac{\rho}{\textit{MR}_{\sf it}\left(t_1^{\star}, t_2^{\star}\right)}$. Then $\textmd{OUT}
\left(
{\textit{DQ}}_{{\sf mr}, {\sf ts}}
\right)$ and $\textmd{OUT}_{{\sf mr}, {\sf ts}}^{\sf opt}$ can be rewritten as
\begin{align}
\textmd{OUT}
\left(
{\textit{DQ}}_{{\sf mr}, {\sf ts}}
\right)
& =
\underbrace{\textmd{Prob}
\left\{
{\bf H} \in \mathcal{H}_1,
{\textit{DQ}}_{{\sf mr}, {\sf ts}} \left({\bf H}\right) < \rho
\right\}}_{ = \textmd{OUT}_1}
\nonumber\\
& +
\underbrace{\textmd{Prob}
\left\{
{\bf H} \in \mathcal{H}_2,
{\textit{DQ}}_{{\sf mr}, {\sf ts}} \left({\bf H}\right) < \rho
\right\}}_{ = \textmd{OUT}_2}
\nonumber\\
& +
\underbrace{\textmd{Prob}
\left\{
{\bf H} \in \mathcal{H}_3,
{\textit{DQ}}_{{\sf mr}, {\sf ts}} \left({\bf H}\right) < \rho
\right\}}_{ = \textmd{OUT}_3},
\nonumber\\
\textmd{OUT}_{{\sf mr}, {\sf ts}}^{\sf opt}
& =
\textmd{Prob}
\left\{
{\bf H} \in \mathcal{H}_1
\right\}.\nonumber
\end{align}
To prove $\textmd{OUT}
\left(
{\textit{DQ}}_{{\sf mr}, {\sf ts}}
\right) = \textmd{OUT}_{{\sf mr}, {\sf ts}}^{\sf opt}$, it is sufficient to prove $\textmd{OUT}_{{\sf mr}, {\sf ts}}^{\sf opt} = \textmd{OUT}_1$ and $\textmd{OUT}_2 = \textmd{OUT}_3 = 0$.

For any ${\bf H} \in \mathcal{H}_1$, $t_{1, \min} + t_{2, \min} > 1$ is equivalent to $\textit{MR}_{\sf it}\left(t_1^{\star}, t_2^{\star}\right) < \rho$, then ${\bf 1}\left({{\bf H} \in \mathcal{H}_1}\right) = {\bf 1}\left({{\bf H} \in \mathcal{H}_1, \textit{DQ}_{{\sf mr}, {\sf ts}}({\bf H}) < \rho}\right)$. Thus $\textmd{OUT}_1 = \textmd{E}\left[{\bf 1}\left({{\bf H} \in \mathcal{H}_1, \textit{DQ}_{{\sf mr}, {\sf ts}}({\bf H}) < \rho}\right)\right] = \textmd{E}\left[{\bf 1}\left({ {\bf H} \in \mathcal{H}_1}\right)\right] = \textmd{OUT}_{{\sf mr}, {\sf ts}}^{\sf opt}$.

Besides, $\textmd{OUT}_2 \leq \textmd{Prob}\left\{t_{1, \min} + t_{2, \min} = 1\right\} = \textmd{Prob}\left\{\textit{MR}_{\sf it}\left(t_1^{\star}, t_2^{\star}\right) = \rho\right\} = 0$, which is from the fact that the probability of a continuous r.v. assuming a specific value is zero. Since $\textmd{OUT}_2 \geq 0$,  $\textmd{OUT}_2 = 0$.

To prove $\textmd{OUT}_3=0$, it is sufficient to show for any ${\bf H} \in \mathcal{H}_{3}$, ${\textit{MR}_{\sf ts}\left( {\textit{DQ}}_{{\sf mr}, {\sf ts}}\left({\bf H}\right)\right) \geq \rho}$.  Let $t_{k, \min} = \left[0.b_{k, 1}b_{k, 2}\cdots\right]_{2}$.

\begin{lemma}
For any ${\bf H} \in \mathcal{H}_3$, $\textmd{\rm ENC}_{{\sf mr}, {\sf ts}, k}^l \left({\bf h}_k\right) = b_{k, l}$, $t_{k, \min}^{{\sf lb}, l} = \left[0.b_{k, 1} b_{k, 2}\cdots b_{k, l}\right]_{2}$ and $t_{k, \min}^{{\sf ub}, l} = t_{k, \min}^{{\sf lb}, l} + 2^{-l}$ when $k = 1, 2$ and $l \in \mathtt{N} - \{0\}$.
\end{lemma}
The proof of Lemma 1 is given in Appendix C.  Since $t_{1, \min} + t_{2, \min} < 1$, there must exist $\hat{l}\in\mathtt{N}$ such that $t_{1, \min} + t_{2, \min} \leq 1 - 2^{-\hat{l}}$, or equivalently,
 \begin{align}
 \label{Violation}
 \left[0.b_{1, 1}b_{1, 2}\cdots b_{1, \hat{l}}\cdots\right]_{2} + \left[0.b_{2, 1}b_{2, 2}\cdots b_{2, \hat{l}}\cdots\right]_{2} \leq \left[0.\underbrace{11\cdots 1}_{\hat{l}}\right]_{2}.
 \end{align}
 All $(t_{1, \min}, t_{2, \min})$s satisfying \eqref{Violation} can be categorized into the following two types:
 \begin{enumerate}
 \item $\exists 1 \leq l^{'} \leq \hat{l}$ such that $\left(b_{1, l^{'}}, b_{2, l^{'}}\right) = (0, 0)$ and $\left(b_{1, l}, b_{2, l}\right) \in \left\{(0, 1), (1, 0)\right\}$ for $l = 1, \ldots, l^{'} - 1$;
 \item $\left(b_{1, l}, b_{2, l}\right) \in \left\{(0, 1), (1, 0)\right\}$ for any $l\leq \hat{l}$ and $\left(b_{1, \hat{l} + 1}, b_{2, \hat{l} + 1}\right) =(0, 0)$.
 \end{enumerate}

 For 1), by Lemma 1, $\textmd{\rm ENC}_{{\sf mr}, {\sf ts}, 1}^{{l}^{'}} \left({\bf h}_k\right) = \textmd{\rm ENC}_{{\sf mr}, {\sf ts}, 2}^{{l}^{'}} \left({\bf h}_k\right) = 0$, then the distributed quantization process will stop at round $l^{'}$ and
\begin{align}
\textit{DQ}_{{\sf mr}, {\sf ts}}\left({\bf H}\right) &= \left(\frac{t_{1, \min}^{{\sf lb}, l^{'} - 1} + t_{1, \min}^{{\sf ub}, l^{'} - 1}}{2}, \frac{t_{2, \min}^{{\sf lb}, l^{'} - 1} + t_{2, \min}^{{\sf ub}, l^{'} - 1}}{2}
\right)
\nonumber\\& = \left(\left[0. b_{1, 1}\cdots b_{1, {l^{'}-1}} 1\right]_{2}, \left[0. b_{2, 1}\cdots b_{2, {l^{'}-1}} 1\right]_{2}
\right).\nonumber
\end{align}
 Since $t_{k, \min} \leq \left[0. b_{k, 1}\cdots b_{k, {l^{'}-1}} 1\right]_{2}$, ${\textit{MR}_{\sf ts}\left( {\textit{DQ}}_{{\sf mr}, {\sf ts}}\left({\bf H}\right)\right) \geq \rho}$.

 For 2),  by Lemma 1, $\textmd{\rm ENC}_{{\sf mr}, {\sf ts}, 1}^{\hat{l} + 1} \left({\bf h}_k\right) = \textmd{\rm ENC}_{{\sf mr}, {\sf ts}, 2}^{\hat{l} + 1} \left({\bf h}_k\right) = 0$, then the distributed quantization process will stop at round $\hat{l} + 1$ and
\begin{align}
\textit{DQ}_{{\sf mr}, {\sf ts}}\left({\bf H}\right) &= \left(\frac{t_{1, \min}^{{\sf lb}, \hat{l}} + t_{1, \min}^{{\sf ub}, \hat{l}}}{2}, \frac{t_{2, \min}^{{\sf lb}, \hat{l}} + t_{2, \min}^{{\sf ub}, \hat{l}}}{2}
\right)
\nonumber\\& = \left(\left[0. b_{1, 1}\cdots b_{1, \hat{l}} 1\right]_{2}, \left[0. b_{2, 1}\cdots b_{2, \hat{l}} 1\right]_{2}
\right).\nonumber
\end{align}
 Since $t_{k, \min} \leq \left[0. b_{k, 1}\cdots b_{k, {l^{'}}} 1\right]_{2}$, ${\textit{MR}_{\sf ts}\left( {\textit{DQ}}_{{\sf mr}, {\sf ts}}\left({\bf H}\right)\right) \geq \rho}$. Therefore, for any ${\bf H} \in \mathcal{H}_3$, ${\textit{MR}_{\sf ts}\left( {\textit{DQ}}_{{\sf mr}, {\sf ts}}\left({\bf H}\right)\right) \geq \rho}$ and $\textmd{OUT}_3 = 0$. To summarize, $\textmd{OUT}
\left(
{\textit{DQ}}_{{\sf mr}, {\sf ts}}
\right) = \textmd{OUT}_{{\sf mr}, {\sf ts}}^{\sf opt}$.

Now, let's prove the upper bound given in \eqref{FR_Time_Sharing}.  Let
\begin{align}
{\mathcal{{R}}}_l = \left\{{\bf H}: \text{ the quantization process of } {\textit{DQ}}_{{\sf mr}, {\sf ts}} \text{ will stop after round } l \right\},\nonumber
\end{align}
 for $l \in \mathtt{N}$. From Lemma 1 and the description of $\textit{DQ}_{{\sf mr}, {\sf ts}}$, for $l \geq 1$,
 \begin{align}
 \mathcal{R}_l = \left\{{\bf H}: (b_{1, l}, b_{2, l}) = (0, 0) \text{ or } (1, 1), (b_{1, m}, b_{2, m}) \in\{(0, 1), (1, 0)\}, m = 1, \ldots, l - 1\right\}.\nonumber
 \end{align}
More specifically, ${\mathcal{{R}}}_l = \bigcup_{q = 0}^{2^{l} - 1} \left\{{\mathcal{{R}}}_{l, q}^{(1)}\cup{\mathcal{{R}}}_{l, q}^{(2)}\right\}$, where
\begin{align}
\label{DCP_1}
\begin{array}{l}
\mathcal{R}_{l, q}^{(1)}
= \left\{{\bf H}: \frac{2q}{2^{l}} \leq t_{1, \min} \leq \frac{2q+1}{2^{l}}, 1 - \frac{2q+2}{2^{l}} \leq t_{2, \min}\leq 1-\frac{2q+1}{2^{l}}, 0 < t_{1, \min}, t_{2, \min} < 1\right\}
\\
\hspace{8.5mm}-\left\{{\bf H}: t_{1, \min} = \frac{2q+1}{2^{l}}, t_{2, \min} = 1-\frac{2q+1}{2^{l}}\right\},
\\
\mathcal{R}_{l, q}^{(2)}
= \left\{{\bf H}: \frac{2q + 1}{2^{l}} \leq t_{1, \min} \leq \frac{2q+2}{2^{l}}, 1 - \frac{2q+1}{2^{l}} \leq t_{2, \min} \leq 1 - \frac{2q}{2^{l}}, 0 < t_{1, \min}, t_{2, \min} < 1\right\}
\\
\hspace{8.5mm}- \left\{{\bf H}: t_{1, \min} = \frac{2q+1}{2^{l}}, t_{2, \min} = 1-\frac{2q+1}{2^{l}}\right\}.
\end{array}
\end{align}
It follows from \eqref{DCP_1}  that
\begin{align}
\label{Compound_R_W}
\begin{array}{l}
\bigcup_{w = l}^{\infty} {\mathcal{R}}_{w} \subseteq  \left\{\bigcup_{u=0}^{2^{l-1}-1}\left\{{\bf H}: \frac{1}{2} - \frac{u + 1}{2^{l}}  \leq  t_{1, \min} \leq \frac{1}{2} - \frac{u}{2^l}, \frac{1}{2} + \frac{u }{2^{l}} \leq t_{2, \min}\leq \frac{1}{2} + \frac{u + 1}{2^l}\right\}\right. \\
  \left.
  \hspace{21.5mm}\cup
  \bigcup_{u=0}^{2^{l-1}-1}\left\{{\bf H}: \frac{1}{2} + \frac{u }{2^{l}} \leq t_{1, \min}\leq \frac{1}{2} + \frac{u + 1}{2^l}, \frac{1}{2} - \frac{u + 1}{2^{l}} \leq  t_{2, \min} \leq \frac{1}{2} - \frac{u}{2^l}\right\}
  \right\}.
\end{array}
\end{align}
 Since $2(l+1)$ bits are fed back in total after round $l$, the average feedback rate is given as
\begin{align}
\label{DQ_FR_Temp}
\textmd{FR}
\left({\textit{DQ}}_{{\sf mr}, {\sf ts}}\right) & = \sum_{l = 0}^{\infty} 2(l+1)\textmd{Prob}\left\{{\bf H} \in {\mathcal{{R}}}_l\right\},\nonumber
\\
& =
2 \textmd{Prob}\left\{{\bf H} \in {\mathcal{{R}}}_0\right\}
+
4 \textmd{Prob}\left\{{\bf H} \in {\mathcal{{R}}}_1\right\}
+
\sum_{l = 2}^{\infty} (2l+2) \textmd{Prob}\left\{{\bf H} \in {\mathcal{{R}}}_l \right\}
\nonumber\\
& =
2
+
2 \textmd{Prob}\left\{{\bf H} \in {\mathcal{{R}}}_1\right\}
+
2 \sum_{l = 2}^{\infty} l \times\textmd{Prob}\left\{{\bf H} \in {\mathcal{{R}}}_l \right\}
\nonumber\\
&\leq 2
+
2 \textmd{Prob}\left\{{\bf H} \in {\mathcal{{R}}}_1\right\}
+
2\sum_{l = 2}^{\infty}l \times \textmd{Prob}\left\{{\bf H} \in \bigcup_{w = l}^{\infty} {\mathcal{{R}}}_w \right\}.
\end{align}
It is trivial to obtain the PDF of $t_{k, \min}$ as $f_{t_{k, \min}}(x) = \frac{\rho \log 2 }{P x^2 } e^{-\frac{e^{\frac{\rho \log 2}{x} - 1}}{P}} e^{\frac{\rho \log 2}{x}}, x > 0$ for $k = 1, 2$.  Since  ${\mathcal{{R}}}_1 \subseteq \left\{{\bf H}: 0 \leq t_{1, \min}, t_{2, \min} \leq \frac{1}{2} \text{ or } \frac{1}{2} \leq t_{1, \min}, t_{2, \min} \leq 1 \right\}$, the upper bound on $\textmd{Prob}\left\{{\bf H} \in {\mathcal{{R}}}_1\right\}$ is derived as
\begin{align}
\label{Temp_Prob}
\textmd{Prob}\left\{{\bf H} \in {\mathcal{{R}}}_1\right\}
&  \leq  \int_0^{\frac{1}{2}} f_{t_{1, \min}}(x_1) {\rm d}x_1\int_0^{\frac{1}{2}} f_{t_{2, \min}}(x_2) {\rm d}x_2
  +
  \int_{\frac{1}{2}}^1 f_{t_{1, \min}}(x_1) {\rm d}x_1\int_{\frac{1}{2}}^1 f_{t_{2, \min}}(x_2) {\rm d}x_2
  \nonumber\\
  & \leq \int_0^{1} f_{t_{1, \min}}(x_1) {\rm d}x_1 = e^{-\frac{e^{\rho\log 2}-1}{P}} \leq e^{-\frac{{\rho\log 2}}{P}},
\end{align}
where the inequalities arise from $\int_0^{\frac{1}{2}}
f_{t_{2, \min}}(x_2){\rm d}x_2 \leq 1$, $\int_{\frac{1}{2}}^1
f_{t_{2, \min}}(x_2){\rm d}x_2 \leq 1$, and $e^x - 1 \geq x$ for $x\geq 0$.

When $l \geq 2$, from \eqref{Compound_R_W},  $\textmd{Prob}\left\{{\bf H} \in \bigcup_{w = l}^{\infty} {\mathcal{{R}}}_w \right\}$ can be bounded by
\begin{align}
\textmd{Prob}
\left\{
{\bf H} \in \bigcup_{w = l}^{\infty}{\mathcal{R}}_{w}
\right\}
& \leq
\sum_{u = 0}^{2^{l-1}-1}
\int_{\frac{1}{2} - \frac{u + 1}{2^{l}}}^{\frac{1}{2} - \frac{u}{2^l}}
f_{t_{1, \min}}(x_1){\rm d}x_1
\int_{\frac{1}{2} + \frac{u }{2^{l}}}^{\frac{1}{2} + \frac{u + 1}{2^l}}
f_{t_{2, \min}}(x_2){\rm d}x_2
\nonumber\\
& +
\sum_{u = 0}^{2^{l-1}-1}
\int_{\frac{1}{2} + \frac{u }{2^{l}}}^{\frac{1}{2} + \frac{u + 1}{2^l}}
f_{t_{1, \min}}(x_1){\rm d}x_1
\int_{\frac{1}{2} - \frac{u + 1}{2^{l}}}^{\frac{1}{2} - \frac{u}{2^l}}
f_{t_{2, \min}}(x_2){\rm d}x_2
\nonumber\\
& =
2\sum_{u = 0}^{2^{l-1}-1}
\int_{\frac{1}{2} - \frac{u + 1}{2^{l}}}^{\frac{1}{2} - \frac{u}{2^l}}
f_{t_{1, \min}}(x_1){\rm d}x_1
\int_{\frac{1}{2} + \frac{u }{2^{l}}}^{\frac{1}{2} + \frac{u + 1}{2^l}}
f_{t_{2, \min}}(x_2){\rm d}x_2.
\nonumber
\end{align}
When $\frac{1}{2} + \frac{u }{2^{l}} \leq x_2 \leq  \frac{1}{2} + \frac{u + 1}{2^l}$, $\frac{1}{2}\leq x_2 \leq 1$, thus $f_{t_{2, \min}}(x_2) = \frac{\rho \log 2 }{P x_2^2 } e^{-\frac{e^{\frac{\rho \log 2}{x_2} - 1}}{P}} e^{\frac{\rho \log 2}{x_2}} \leq \frac{4 \rho \log 2 }{P } e^{-\frac{e^{{\rho \log 2} - 1}}{P}} e^{{2 \rho \log 2}}$. Then the upper bound on $\textmd{Prob}\left\{{\bf H} \in \bigcup_{w = l}^{\infty} {\mathcal{{R}}}_w \right\}$ is further derived as
\begin{align}
\label{Temp_Tail}
\textmd{Prob}
\left\{
{\bf H} \in \bigcup_{w = l}^{\infty}{\mathcal{R}}_{w}
\right\}
& \leq
\frac{8\rho \log 2}{P}\sum_{u = 0}^{2^{l-1}-1}
\int_{\frac{1}{2} - \frac{u + 1}{2^{l}}}^{\frac{1}{2} - \frac{u}{2^l}}
f_{t_{1, \min}}(x_1) {\rm d}x_1
\int_{\frac{1}{2} + \frac{u }{2^{l}}}^{\frac{1}{2} + \frac{u + 1}{2^l}}
{e^{-\frac{e^{{\rho}{\log 2} - 1}}{P}} e^{{2\rho}{\log 2}}} {\rm d}x_2
\nonumber\\
& \leq
\frac{8\rho \log 2}{P}\sum_{u = 0}^{2^{l-1}-1}
\int_{\frac{1}{2} - \frac{u + 1}{2^{l}}}^{\frac{1}{2} - \frac{u}{2^l}}
f_{t_{1, \min}}(x_1) {\rm d}x_1
\int_{\frac{1}{2} + \frac{u }{2^{l}}}^{\frac{1}{2} + \frac{u + 1}{2^l}}
{e^{-\frac{{{\rho}{\log 2}}}{P}} e^{{2\rho}{\log 2}}} {\rm d}x_2
\nonumber\\
& =
{8\rho e^{{2\rho}{\log 2}}}{\log 2}\times \frac{e^{-\frac{\rho \log 2}{P}}}{P}
\times \frac{1}{2^l}\sum_{u = 0}^{2^{l-1}-1}
\int_{\frac{1}{2} - \frac{u + 1}{2^{l}}}^{\frac{1}{2} - \frac{u}{2^l}}
f_{t_{1, \min}}(x_1) {\rm d}x_1
\nonumber\\
& =
{8\rho e^{{2\rho}{\log 2}}}{\log 2}\times \frac{e^{-\frac{\rho \log 2}{P}}}{P}
\times \frac{1}{2^l} \int_0^{\frac{1}{2}} f_{t_{1, \min}}(x_1) {\rm d}x_1
\nonumber\\
& \leq
{8\rho e^{{2\rho}{\log 2}}}{\log 2}\times \frac{e^{-\frac{\rho \log 2}{P}}}{P}
\times \frac{1}{2^l}.
\end{align}
Subsituting \eqref{Temp_Prob}, \eqref{Temp_Tail} into \eqref{DQ_FR_Temp} and using the fact that $\sum_{l = 2}^{\infty}\frac{l}{2^l}$ is finite yield the upper bound in \eqref{FR_Time_Sharing}.

\end{IEEEproof}

\section{Proof of Lemma 1}
\begin{IEEEproof}
Based on the procedures in $\textit{DQ}_{{\sf mr}, {\sf ts}}$, $t_{k, \min}^{{\sf lb}, l} \leq t_{k, \min} \leq t_{k, \min}^{{\sf ub}, l}$ for $l \in \mathtt{N} - \{0\}$.

It is straightforward to verify Lemma 1 holds when $l = 1$. By induction, assume Lemma 1 holds when $l \leq m$ where $m \geq 2$. For $l = m + 1$,\footnote{We assume the quantization process in $\textit{DQ}_{{\sf mr}, {\sf ts}}$ still continues in round $m + 1$. Otherwise, it is not necessary to consider Lemma 1 when $l = m + 1$.} according to $\textit{DQ}_{{\sf mr}, {\sf ts}}$, $\textmd{\rm ENC}_{{\sf mr}, {\sf ts}, k}^{m + 1} \left({\bf h}_k\right) = {\bf 1}\left({t_{k, \min} \geq \frac{t_{k, \min}^{{\sf lb}, m} + t_{k, \min}^{{\sf ub}, m}}{2}}\right)$, and
\begin{align}
\frac{t_{k, \min}^{{\sf lb}, m} + t_{k, \min}^{{\sf ub}, m}}{2} = \left[0.b_{k, 1} b_{k, 2}\cdots b_{k, m}\right]_{2} + 2^{-m-1} = \left[0.b_{k, 1} b_{k, 2}\cdots b_{k, m} 1\right]_{2}.\nonumber
\end{align}

 If ${t_{k, \min} \geq \left[0.b_{k, 1} b_{k, 2}\cdots b_{k, m} 1\right]_{2} =\frac{t_{k, \min}^{{\sf lb}, m} + t_{k, \min}^{{\sf ub}, m}}{2}}$, it must have $b_{k, m + 1} = 1 = \textmd{\rm ENC}_{{\sf mr}, {\sf ts}, k}^{m + 1} \left({\bf h}_k\right)$. Then $t_{k, \min}^{{\sf lb}, m + 1} = \frac{t_{k, \min}^{{\sf lb}, m} + t_{k, \min}^{{\sf ub}, m}}{2} = \left[0.b_{k, 1} b_{k, 2}\cdots b_{k, m} b_{k, m + 1}\right]_{2}$ and $t_{k, \min}^{{\sf ub}, m + 1} = t_{k, \min}^{{\sf ub}, m }  = t_{k, \min}^{{\sf lb}, m } + 2^{-m} = t_{k, \min}^{{\sf lb}, m+1 } + 2^{-m - 1}$.

  If ${t_{k, \min} < \left[0.b_{k, 1} b_{k, 2}\cdots b_{k, m} 1\right]_{2} =\frac{t_{k, \min}^{{\sf lb}, m} + t_{k, \min}^{{\sf ub}, m}}{2}}$, since $t_{k, \min}\geq t_{k, \min}^{{\sf lb}, m} = \left[0.b_{k, 1} b_{k, 2}\cdots b_{k, m}\right]_{2}$,  it must have $b_{k, m + 1} = 0 = \textmd{\rm ENC}_{{\sf mr}, {\sf ts}, k}^{m + 1} \left({\bf h}_k\right)$. Then $t_{k, \min}^{{\sf lb}, m + 1} = t_{k, \min}^{{\sf lb}, m }  =\left[0.b_{k, 1} b_{k, 2}\cdots b_{k, m} 0\right]_{2} = \left[0.b_{k, 1} b_{k, 2}\cdots b_{k, m} b_{k, m + 1}\right]_{2}$ and $t_{k, \min}^{{\sf ub}, m + 1} = \frac{t_{k, \min}^{{\sf lb}, m} + t_{k, \min}^{{\sf ub}, m}}{2} = \left[0.b_{k, 1} b_{k, 2}\cdots b_{k, m} 1\right]_{2} =t_{k, \min}^{{\sf lb}, m + 1} + 2^{-m-1} $. Therefore, Lemma 1 holds when $l = m + 1$. In conclusion, Lemma 1 holds for any $l \in \mathtt{N} - \{0\}$.

\end{IEEEproof}

\section{Proof of Theorem 3}
\begin{IEEEproof}
For a given $M \in \mathtt{N} - \{0\}$, define a global quantizer which selects the interference transmission pair that maximizes $\textit{MR}_{\sf it}\left(p_1, p_2\right)$ among the codebook $\mathcal{C}_{{\sf unif}}$  as
\begin{align}
\textit{GQ}_{{\sf mr}, {\sf it}}\left({\bf H}\right)
=
\argmax\limits_{(p_1, p_2) \in \mathcal{C}_{{\sf unif}}}
\textit{MR}_{\sf it}\left(p_1, p_2\right),\nonumber
\end{align}
where $\mathcal{C}_{{\sf unif}} = \left\{(1, 1), (1, \frac{m}{M}), (\frac{m}{M}, 1):  m = 1, \ldots, M - 1\right\}$.

Let $\textmd{OUT}
\left(\textit{GQ}_{{\sf mr}, {\sf it}}\right)
=
\textmd{Prob}
\left\{
\textit{MR}_{{\sf it}}\left(\textit{GQ}_{{\sf mr}, {\sf it}}\left({\bf H}\right) \right)< \rho
\right\}$. First, let us show that $\textmd{OUT}
\left({\textit{DQ}}_{{\sf mr}, {\sf it}}\right) = \textmd{OUT}
\left(\textit{GQ}_{{\sf mr}, {\sf it}}\right)$.

According to ${\textit{GQ}}_{{\sf mr}, {\sf it}}$,  an outage event happens if and only if $\textit{MR}_{{\sf it}}\left(p_1, p_2\right) < {\rho}$ for any $(p_1, p_2) \in \mathcal{C}_{{\sf unif}}$. In  ${\textit{DQ}}_{{\sf mr}, {\sf it}}$, an outage occurs if and only if the following conditions are satisfied: (i) receiver 2 sends ``0'' after round $0$; (ii) $\log_2\left(1 + \frac{\textmd{ENC}_{{\sf mr}, {\sf it}, 2}^{1} \left({\bf h}_2\right) P H_{1, 1}}{ P H_{2, 1} + 1}\right) < \rho$. (i) happens because $\log_2\left(1 + \frac{ \textmd{ENC}_{{\sf mr}, {\sf it}, 1}^{0} \left({\bf h}_1\right) P H_{2, 2}}{ P H_{1, 2} + 1}\right) < \rho$. It means for $x \in\mathcal{C}_{M}$, $\log_2\left(1 + \frac{ P H_{1, 1}}{ x P H_{2, 1} + 1}\right) \geq \rho$ and $\log_2\left(1 + \frac{ x P H_{2, 2}}{ P H_{1, 2} + 1}\right) \geq \rho$ cannot hold simultaneously, or equivalently,  ${{\textit{MR}}}_{\sf it}\left(p_1, p_2\right) < {\rho}$ for $(1, p_2) \in \mathcal{C}_{{\sf unif}}$. Similarly, (ii) means $\log_2\left(1 + \frac{ x P H_{1, 1}}{  P H_{2, 1} + 1}\right) \geq \rho$ and $\log_2\left(1 + \frac{  P H_{2, 2}}{ x P H_{1, 2} + 1}\right) \geq \rho$ cannot stand at the same time for $x\in\mathcal{C}_{M}$, which is to say, ${{\textit{MR}}}_{\sf it}\left(p_1, p_2\right) < {\rho}$ for $(p_1, 1) \in \mathcal{C}_{\textmd{unif}}$. Thus, (i) and (ii) both happen means ${{\textit{MR}}}_{\sf it}\left(p_1, p_2\right) < {\rho}$ for any $(p_1, p_2)\in \mathcal{C}_{\textmd{unif}}$. i.e.,  ${\bf 1}\left({\textit{MR}_{{\sf it}}\left(\textit{GQ}_{{\sf mr}, {\sf it}}\left({\bf H}\right) \right)< \rho}\right)={\bf 1}\left({\textit{MR}_{\sf it}\left( {\textit{DQ}}_{{\sf mr}, {\sf it}}\left({\bf H}\right)\right) < \rho}\right)$. Hence, we have $\textmd{OUT}
\left({\textit{DQ}}_{{\sf mr}, {\sf it}}\right)= \textmd{OUT}
\left(\textit{GQ}_{{\sf mr}, {\sf it}}\right)$ since $\textmd{OUT}
\left({\textit{DQ}}_{{\sf mr}, {\sf it}}\right)
 =
\textmd{E}\left[{\bf 1}\left({\textit{MR}_{\sf it}\left( {\textit{DQ}}_{{\sf mr}, {\sf it}}\left({\bf H}\right)\right) < \rho}\right)\right]$ and $\textmd{OUT}
\left(\textit{GQ}_{{\sf mr}, {\sf it}}\right)
 =
\textmd{E}\left[{\bf 1}\left({\textit{MR}_{{\sf it}}\left(\textit{GQ}_{{\sf mr}, {\sf it}}\left({\bf H}\right) \right)< \rho}\right)\right]$.

To prove \eqref{DQ_OUT_OUT}, it is sufficient to show $\textmd{OUT}
\left(\textit{GQ}_{{\sf mr}, {\sf it}}\right) \leq \textmd{OUT}_{{\sf mr}, {\sf it}}^{\sf opt} + \frac{C_1}{M}$. Define another quantizer $\tilde{\textit{GQ}}_{{\sf mr}, {\sf it}}$ that selects the interference transmission pair according to
\begin{align}
\label{OUT_Sub_TPV}
\tilde{\textit{GQ}}_{{\sf mr}, {\sf it}}\left({\bf H}\right) =
\left\{
\begin{matrix}
\left(\hat{p}_1, 1
\right), & \frac{{H}_{1, 1}}{ {H}_{2, 1} + \frac{1}{P}} \geq \frac{{H}_{2, 2}}{ {H}_{1, 2} + \frac{1}{P}},\\
\left(1, \hat{p}_2
\right), & \frac{{H}_{1, 1}}{ {H}_{2, 1} + \frac{1}{P} } < \frac{{H}_{2, 2}}{ {H}_{1, 2} + \frac{1}{P}},
\end{matrix}
\right.
\end{align}
where
\begin{align}
\label{hat_p}
\hat{p}_1   =
\max\limits_{
\begin{subarray}{c}
x\in \mathcal{C}_M, x\leq {p}_1^{\star}
\end{subarray}} x, \hat{p}_2   =
\max\limits_{
\begin{subarray}{c}
x\in\mathcal{C}_M, x\leq {p}_2^{\star}
\end{subarray}} x.
\end{align}
The network outage probability of minimum rate achieved by $\tilde{\textit{GQ}}_{{\sf mr}, {\sf it}}$  is $\textmd{OUT}
\left(\tilde{\textit{GQ}}_{{\sf mr}, {\sf it}}\right)
=
\mathtt{Prob}
\left\{
\tilde{\textit{GQ}}_{{\sf mr}, {\sf it}}\left({\bf H}\right) < \rho
\right\}$. Since ${\textit{GQ}}_{{\sf mr}, {\sf it}}\left({\bf H}\right) \geq \tilde{\textit{GQ}}_{{\sf mr}, {\sf it}}\left({\bf H}\right)$,  $\textmd{OUT}
\left(\textit{GQ}_{{\sf mr}, {\sf it}}\right)  \leq \textmd{OUT}
\left(\tilde{\textit{GQ}}_{{\sf mr}, {\sf it}}\right)$. Hence, to prove \eqref{DQ_OUT_OUT}, it is sufficient to prove $\textmd{OUT}
\left(\tilde{\textit{GQ}}_{{\sf mr}, {\sf it}}\right) - \textmd{OUT}_{{\sf mr}, {\sf it}}^{\sf opt} \leq \frac{C_1}{M}$.

Let $\bar{\rho} = 2^{\rho}-1$, $H_{121} = \frac{{H}_{1, 1}} {{H}_{2, 1} + \frac{1}{P}}$, $H_{212} = \frac{{H}_{2, 2}} {{H}_{1, 2} + \frac{1}{P}}$, and $\alpha = \frac{1}{M}$. When $M = 1$, $\textmd{OUT}
\left(\tilde{\textit{GQ}}_{{\sf mr}, {\sf it}}\right) = \textmd{Prob}\left\{\textit{MR}_{\sf it}(1, 1) < \rho\right\}$. Let $C_2 = \textmd{Prob}\left\{\textit{MR}_{\sf it}(1, 1) < \rho\right\}$, then $\textmd{OUT}
\left(\tilde{\textit{GQ}}_{{\sf mr}, {\sf it}}\right)  \leq \frac{C_2}{M}$. When $M\geq 1$, $0 < \alpha \leq \frac{1}{2} < 1$. $\textmd{OUT}_{{\sf mr}, {\sf it}}^{\sf opt}$ and $\textmd{OUT}
\left(\tilde{\textit{GQ}}_{{\sf mr}, {\sf it}}\right)$ are rewritten as
\begin{align}
\textmd{OUT}_{{\sf mr}, {\sf it}}^{\sf opt}&
 = \textmd{Prob}\left\{H_{121} \geq H_{212}, p_1^{\star} H_{121} <\bar{\rho}\right\} + \textmd{Prob}\left\{H_{121} < H_{212}, p_2^{\star} H_{212} < \bar{\rho}\right\}, \nonumber\\
 \textmd{OUT}
\left(\tilde{\textit{GQ}}_{{\sf mr}, {\sf it}}\right)
&  = {\textmd{Prob}
\left\{
H_{121} \geq H_{212},
\hat{p}_1 H_{121} < \bar{\rho}
\right\}}
  +
{\textmd{Prob}
\left\{
H_{121} < H_{212},
\hat{p}_2 H_{212} < \bar{\rho}
\right\}}, \nonumber
\end{align}
 then $\textmd{OUT}
\left(\tilde{\textit{GQ}}_{{\sf mr}, {\sf it}}\right) - \textmd{OUT}_{{\sf mr}, {\sf it}}^{\sf opt}$ is derived as
\begin{align}
\label{OUT_UB_1}
& \textmd{OUT}
 \left(\tilde{\textit{GQ}}_{{\sf mr}, {\sf it}}\right) - \textmd{OUT}_{{\sf mr}, {\sf it}}^{\sf opt}
\nonumber\\
& =
{\textmd{Prob}
\left\{
H_{121} \geq H_{212}, p_1^{\star} H_{121} \geq \bar{\rho},
\hat{p}_1 H_{121} < \bar{\rho}
\right\}}
\nonumber\\
&  +
{\textmd{Prob}
\left\{
H_{121} < H_{212},
p_2^{\star} H_{212} \geq \bar{\rho},
\hat{p}_2 H_{212} < \bar{\rho}
\right\}}\nonumber\\
& =
2 {\textmd{Prob}
\left\{
H_{121} \geq H_{212},
p_1^{\star} H_{121} \geq \bar{\rho},
\hat{p}_1 H_{121} < \bar{\rho}
\right\}}
\nonumber\\
& \leq
2 {\textmd{Prob}
\left\{
H_{121} \geq H_{212}, p_1^{\star} H_{121} \geq \bar{\rho},
\left(p_1^{\star} - \alpha\right) H_{121} < \bar{\rho}
\right\}}
\nonumber\\
& =
2 {\textmd{Prob}
\left\{
H_{121} \geq H_{212},
\frac{\bar{\rho}}{H_{121}} \leq p_1^{\star} < \frac{\bar{\rho}}{H_{121}} + \alpha
\right\}},
\end{align}
where the first inequality is from ${p}_1^{\star} - \hat{p}_1 \leq \alpha$ by \eqref{hat_p}. Let $A = \frac{\bar{\rho}}{H_{121}}$ and $B = A + \alpha$. The PDFs of $H_{k, l}$ are $f_{H_{1, 1}}(x) = f_{H_{2, 2}}(x) = e^{-x}$ and $f_{H_{1, 2}}(x) = f_{H_{2, 1}}(x) = \frac{1}{\epsilon}e^{-\frac{x}{\epsilon}}$, $x > 0$, for $k, l = 1, 2$.  Then the PDFs of $H_{121}$ and $H_{212}$ are easily obtained as $f_{H_{121}}(x) = f_{H_{212}}(x) = \frac{e^{-\frac{x}{P}}}{P(\epsilon x + 1)} + \frac{\epsilon e^{-\frac{x}{P}}}{(\epsilon x + 1)^2}$, $x > 0$. From \eqref{First_p}, ${p}_1^{\star}$ is rewritten as ${p}_1^{\star}
=
\frac
{\sqrt{\frac{4P^2 }{H_{121}} {H}_{2, 2}{H}_{1, 2} + 1} - 1}
{2P {H}_{1, 2}}$. Since $0\leq p_1^{\star}\leq 1$, it follows that
\begin{align}
\label{I_1_OUT}
  & \textmd{OUT}
 \left(\tilde{\textit{GQ}}_{{\sf mr}, {\sf it}}\right) - \textmd{OUT}_{{\sf mr}, {\sf it}}^{\sf opt}
\nonumber\\
& \leq
2\textmd{Prob}
\left\{
H_{121} \geq H_{212}, A \leq 1, B > 1, A \leq p_1^{\star}
\right\}
+
2\textmd{Prob}
\left\{
H_{121} \geq H_{212}, B \leq 1,
A
\leq
p_{1}^{\star}
<
B
\right\}
\nonumber\\
& \leq
2\textmd{Prob}
\left\{
H_{121} \geq H_{212}, \bar{\rho} \leq H_{121}  < \frac{\bar{\rho}}{1 - \alpha},
A
\leq
\frac
{\sqrt{\frac{4P^2 }{H_{121}} {H}_{2, 2}{H}_{1, 2} + 1} - 1}
{2P {H}_{1, 2}}
\right\}
\nonumber\\
& +
2\textmd{Prob}
\left\{
H_{121} \geq H_{212},  H_{121} \geq \frac{\bar{\rho}}{1 - \alpha},
A
\leq
\frac
{\sqrt{\frac{4P^2 }{H_{121}} {H}_{2, 2}{H}_{1, 2} + 1} - 1}
{2P {H}_{1, 2}}
<
B
\right\}
\nonumber\\
& \leq
2\underbrace{\textmd{Prob}
\left\{
\bar{\rho} \leq H_{121}  < \frac{\bar{\rho}}{1 - \alpha},
H_{121} A^2 {H}_{1, 2} + \frac{A}{P}H_{121}
\leq
{H}_{2, 2} < H_{121}   {H}_{1, 2} + \frac{H_{121} }{P}
\right\}}_{ = I_{1}}
\nonumber\\
& +
2\underbrace{\textmd{Prob}
\left\{
H_{121}  \geq \frac{\bar{\rho}}{1 - \alpha},
H_{121}  A^2 {H}_{1, 2} + \frac{A}{P}H_{121}
\leq
{H}_{2, 2} < H_{121}  B^2 {H}_{1, 2} + \frac{B}{P} H_{121}
\right\}.}_{= I_{2}}
\end{align}
The upper bound on $I_{1}$ can be derived as
\begin{align}
I_{1}
 \leq
\textmd{Prob}
\left\{
\bar{\rho} \leq H_{121} \leq \frac{\bar{\rho} }{1 - \alpha}
\right\}
=
\int_{\bar{\rho}}^{\frac{\bar{\rho} }{1 - \alpha}}
f_{H_{121}}(x){\rm d}x
 =
\frac
{e^{-\frac{\bar{\rho}}{P}}}
{\epsilon \bar{\rho} + 1}
\left(
1 -
\frac
{\epsilon \bar{\rho} + 1}
{\epsilon \frac{\bar{\rho}}{1 - \alpha} + 1}
e^{-\frac{\bar{\rho}}{P (1 - \alpha)}\alpha}
\right).\nonumber
\end{align}
Since $1 - x e^{-y} \leq 1 - x + xy$ when $0 < x \leq 1, y >0$, $\epsilon \bar{\rho} + 1 \geq 1$, $\frac{1}{1 - \alpha}\geq 1$, and $1 - \alpha \geq \frac{1}{2}$,  $I_{1}$ is further bounded by
\begin{align}
\label{I_11_OUT}
I_{1}
& \leq
\frac
{{e^{-\frac{\bar{\rho}}{P}}}}
{{{\epsilon \bar{\rho} + 1}}}
\left(
1 -
\frac
{\epsilon \bar{\rho} + 1}
{\epsilon \frac{\bar{\rho}}{1 - \alpha} + 1}
+
\frac
{\epsilon \bar{\rho} + 1}
{ \epsilon\frac{\bar{\rho}}{{1 - \alpha}} + 1}
\times
{ \frac{\bar{\rho}}{P {(1 - \alpha)}}\alpha}
\right)
\nonumber\\
& \leq
e^{-\frac{\bar{\rho}}{P}}
\left(
1 -
\frac
{\epsilon \bar{\rho} + 1}
{\epsilon \frac{\bar{\rho}}{1 - \alpha} + 1}
+
\frac
{\epsilon \bar{\rho} + 1}
{ \epsilon \bar{\rho} + 1}
\times
{ \frac{\bar{\rho}}{P \times { \frac{1}{2}}}\alpha}
\right)
\nonumber\\
& \leq
 {{e^{-\frac{\bar{\rho}}{P}}}}
\left[1 + \frac{2\bar{\rho}}{P}\right] \alpha \leq C_3 \alpha,
\end{align}
where $C_3 = 2$. The last inequality arises from $e^{-x}(1 + 2x) \leq 2 e^{-\frac{1}{2}}\leq 2$ for $x \geq 0$. Subsequently, $I_{2}$ is upper-bounded by
\begin{align}
\label{sum_I_2}
I_{2}
& =
\frac{1}{\epsilon}
\int_{\underbrace{\frac{\bar{\rho}}{1 - \alpha}}_{\geq \bar{\rho}}}^{\infty}
f_{H_{121}}(x)
\int_0^{\infty}
e^{-\frac{H_{1, 2}}{\epsilon}}
\int_{x A^2 {H}_{12} + \frac{A}{P}x}^{x B^2 {H}_{12} + \frac{B}{P}x}
e^{-H_{2, 2}}
{\rm d}H_{2, 2}
{\rm d}H_{1, 2}
{\rm d}x
\nonumber\\
& \leq
\frac{1}{\epsilon}
\int_{{\bar{\rho}}}^{\infty}
f_{H_{121}}(x)
\int_0^{\infty}
e^{-\frac{H_{1, 2}}{\epsilon}}
\int_{x A^2 {H}_{12} + \frac{A}{P}x}^{x B^2 {H}_{12} + \frac{B}{P}x}
e^{-H_{2, 2}}
{\rm d}H_{2, 2}
{\rm d}H_{1, 2}
{\rm d}x
\nonumber\\
& =
\frac{1}{\epsilon}
\int_{\bar{\rho}}^{\infty}
f_{H_{121}}(x)
\left(
\frac
{e^{-\frac{A}{P}x}}
{x A^2 + \frac{1}{\epsilon}}
-
\frac
{e^{-\frac{B}{P}x}}
{x B^2 + \frac{1}{\epsilon}}
\right)
{\rm d}x
\nonumber\\
%& =
%\frac{1}{\epsilon}
%\int_{\bar{\rho}}^{\infty}
%f_X(x)
%\frac
%{\frac{e^{-\frac{A}{P}x}}{\epsilon} - \frac{e^{-\frac{B}{P}x}}{\epsilon} + B^2 x  e^{-\frac{A}{P}x} - A^2 x e^{-\frac{B}{P}x}}
%{\left(x A^2 + \frac{1}{\epsilon}\right)\left(x B^2 + \frac{1}{\epsilon}\right)}
%{\rm d}x
%\nonumber\\
& =
 \frac{1}{\epsilon}
\int_{\bar{\rho}}^{\infty}
f_{H_{121}}(x)
\frac
{\overbrace{\left(\frac{e^{-\frac{A}{P}x}}{\epsilon}\right)}^{\leq 1}\overbrace{\left(1 - e^{-\frac{\alpha}{P}x}\right) }^{\leq \frac{\alpha}{P} x}}
{\underbrace{\left(x A^2 + \frac{1}{\epsilon}\right)\left(x B^2 + \frac{1}{\epsilon}\right)}_{\geq \frac{1}{\epsilon^2}}}
{\rm d}x
+
\frac{1}{\epsilon}
\int_{\bar{\rho}}^{\infty}
f_{H_{121}}(x)
\frac
{B^2 x  \overbrace{e^{-\frac{A}{P}x}}^{\leq 1}\overbrace{\left(1 - \frac{A^2}{B^2}e^{-\frac{\alpha}{P}x}\right)}^{\leq 1 - \frac{A^2}{B^2} + \frac{A^2}{B^2} {\frac{\alpha}{P}x}} }
{\left(x A^2 + \frac{1}{\epsilon}\right)\left(x B^2 + \frac{1}{\epsilon}\right)}
{\rm d}x
\nonumber\\
& \leq \frac{1}{\epsilon}
\int_{\bar{\rho}}^{\infty}
f_{H_{121}}(x)
\frac
{\frac{1}{\epsilon}\left(\frac{\alpha}{P} x\right) }
{\frac{1}{\epsilon^2 }}
{\rm d}x
+
\frac{1}{\epsilon}
\int_{\bar{\rho}}^{\infty}
f_{H_{121}}(x)
\frac
{B^2 x  \left(1 - \frac{A^2}{B^2} + \frac{A^2}{B^2} {\frac{\alpha}{P}x}\right) }
{\underbrace{\left(x A^2 + \frac{1}{\epsilon}\right)\left(x B^2 + \frac{1}{\epsilon}\right)}_{\geq \frac{A^2 x}{\epsilon}}}
{\rm d}x
\nonumber\\
&\leq \frac{1}{\epsilon}
\int_{\bar{\rho}}^{\infty}
f_{H_{121}}(x)
\frac
{\frac{1}{\epsilon}\left(\frac{\alpha}{P} x\right) }
{\frac{1}{\epsilon^2 }}
{\rm d}x
+
\frac{1}{\epsilon}
\int_{\bar{\rho}}^{\infty}
f_{H_{121}}(x)
\frac
{B^2 x  \left(1 - \frac{A^2}{B^2} \right) }
{\left(x A^2 + \frac{1}{\epsilon}\right)\left(x B^2 + \frac{1}{\epsilon}\right)}
{\rm d}x
 +
\frac{1}{\epsilon}
\int_{\bar{\rho}}^{\infty}
f_{H_{121}}(x)
\frac
{B^2 x  \left(\frac{A^2}{B^2} {\frac{\alpha}{P}x}\right) }
{ \frac{A^2 x}{\epsilon}}
{\rm d}x
\nonumber\\
& =
\underbrace{\int_{\bar{\rho}}^{\infty}
f_{H_{121}}(x) \frac{2\alpha x}{P} {\rm d}x}_{ = I_{2, 1}}
+
\frac{1}{\epsilon}
\underbrace{\int_{\bar{\rho}}^{\infty}
f_{H_{121}}(x)
\frac
{ (B^2 - A^2 ) x  }
{{\left(x A^2 + \frac{1}{\epsilon}\right)\left(x B^2 + \frac{1}{\epsilon}\right)}}
{\rm d}x.}_{ = I_{2, 2}}
\end{align}
The upper bound on $I_{2, 1}$ is derived as
\begin{align}
\label{I_121_OUT}
I_{2, 1}
& \leq
\int_{\bar{\rho}}^{\infty}
f_{H_{121}}(x) \frac{2\alpha x}{P} {\rm d}x
 =
\frac{2\alpha }{P}
\textmd{E}\left[\frac{H_{1, 1}}{H_{2, 1} + \frac{1}{P}}\right]
 =
\frac{2\alpha }{\epsilon P}
\int_0^{\infty}
e^{-H_{1, 1}}
\int_0^{\infty}
e^{-\frac{H_{2, 1}}{\epsilon}}
\frac{H_{1, 1}}{H_{2, 1} + \frac{1}{P}}
{\rm d}H_{1, 1} {\rm d}H_{2, 1}
\nonumber\\
& =
\frac{2\alpha }{\epsilon P}
\int_0^{\infty}
\frac{e^{-\frac{H_{2, 1}}{\epsilon}}}{H_{2, 1} + \frac{1}{P}}
{\rm d}H_{2, 1}
=
\frac{2\alpha e^{\frac{1}{\epsilon P}}}{\epsilon P}
\int_\frac{1}{\epsilon P}^{\infty}
\frac{e^{-z}}{z}
{\rm d}z
\leq
\frac{2\log(1 + \epsilon P)}{\epsilon P}\alpha  \leq C_4 \alpha,
\end{align}
where $C_4 = 2$. The last inequality is from the exponential integral $\int_x^{\infty}\frac{e^{-y}}{y}{\rm d}y \leq e^{-x} \log\left(1 + \frac{1}{x}\right)$ \cite{Handbook} as well as $\log(1 + x) \leq x$ for $x\geq 0$.

Substituting $A = \frac{\bar{\rho}}{x}$ $ B = A + \alpha$, and $f_{H_{121}}(\cdot)$ into $I_{2, 2}$ yields
\begin{align}
\label{I_122_OUT}
I_{2, 2}
& =
\underbrace{\frac{1}{\epsilon }
\int_{\bar{\rho}}^{\infty}
\left[\frac{ e^{-\frac{x}{P}}}{P (\epsilon x + 1)} + \frac{\epsilon e^{-\frac{x}{P}}}{(\epsilon x + 1)^2}\right]
\frac
{ \alpha^2 x }
{{\left(x A^2 + \frac{1}{\epsilon }\right)\left(x B^2 + \frac{1}{\epsilon }\right)}}
{\rm d}x}_{=I_{2, 2, 1}}
\nonumber\\
& +
\underbrace{\frac{1}{\epsilon }
\int_{{\bar{\rho}}}^{\infty}
\left[\frac{e^{-\frac{x}{P}}}{P (\epsilon x + 1)} + \frac{\epsilon e^{-\frac{x}{P}}}{(\epsilon x + 1)^2}\right]
\frac
{ 2 \alpha \bar{\rho} }
{{\left(x A^2 + \frac{1}{\epsilon }\right)\left(x B^2 + \frac{1}{\epsilon }\right)}}
{\rm d}x}_{=I_{2, 2, 2}}.
\end{align}
$I_{2, 2, 1}$ is bounded by
\begin{align}
\label{Final_I_1221}
I_{2, 2, 1}
& =
\frac{1}{\epsilon }
\int_{\bar{\rho}}^{\infty}
\frac{ {e^{-\frac{x}{P}}}}{P\underbrace{ (\epsilon x + 1)}_{\geq \epsilon x}}
\frac
{ \alpha^2 x }
{\underbrace{\left(x A^2 + \frac{1}{\epsilon }\right)\left(x B^2 + \frac{1}{\epsilon }\right)}_{\geq \frac{1}{\epsilon^2}}}
{\rm d}x
 +
\frac{1}{\epsilon }
\int_{\bar{\rho}}^{\infty}
 \frac{\epsilon \overbrace{e^{-\frac{x}{P}}}^{\leq 1}}{(\epsilon x + 1)^2}
\frac
{ \alpha^2 x }
{{\left(x A^2 + \frac{1}{\epsilon }\right)\left(x B^2 + \frac{1}{\epsilon }\right)}}
{\rm d}x
\nonumber\\
& \leq
\frac{1}{\epsilon }
\int_{0}^{\infty}
\frac{ e^{-\frac{x}{P}}}{P \epsilon x }
\frac
{ \alpha^2 x }
{{\frac{1}{\epsilon^2}}}
{\rm d}x
 +
\frac{1}{\epsilon }
\int_{0}^{\infty}
 \frac{\epsilon }{(\epsilon x + 1)^2}
\frac
{ \alpha^2 x }
{{\left(x A^2 + \frac{1}{\epsilon }\right)\left(x B^2 + \frac{1}{\epsilon }\right)}}
{\rm d}x
\nonumber\\
%& \leq
%{\frac{\alpha^2}{\epsilon }
%\int_{0}^{\infty}
%\frac{e^{-\frac{x}{P}}}{P {\epsilon x}}
%\frac
%{  x }
%{{\frac{1}{\epsilon^2}}}
%{\rm d}x }
%+
%{\frac{\alpha^2}{\epsilon }
%\int_{{ 0}}^{\infty}
%\frac{\epsilon e^{-\frac{x}{P}}}{(\epsilon x + 1)^2}
%\frac
%{  x }
%{{\left(x A^2 + \frac{1}{\epsilon }\right)\left(x B^2 + \frac{1}{\epsilon }\right)}}
%{\rm d}x }
%\nonumber\\
& =
{\alpha^2
\int_{0}^{\infty}
\frac{e^{-\frac{x}{P}}}{P }
{\rm d}x }
+ {\alpha^2
\int_{0}^{\infty}
\frac{1}{(\epsilon x + 1)^2}
\frac
{  x }
{{\left(x A^2 + \frac{1}{\epsilon}\right)\left(x \underbrace{B^2}_{\geq \alpha^2} + \frac{1}{\epsilon}\right)}}
{\rm d}x }
\nonumber\\
& \leq
\alpha ^2 +
\alpha^2
\int_{0}^{\infty}
\frac{1}{(\epsilon x + 1)^2}
\frac
{  x }
{{\left(x \left(\frac{\bar{\rho}}{x}\right)^2 + \frac{1}{\epsilon}\right)\left(x  \alpha ^2 + \frac{1}{\epsilon}\right)}}
{\rm d}x
\nonumber\\
& =
\alpha^2 +
\frac{1}{\epsilon}
\int_{0}^{\infty}
\frac{1}{\left( x + \frac{1}{\epsilon}\right)^2}
\frac
{  x^2 }
{{\left(x + \epsilon \bar{\rho}^2\right)\left(x  + \frac{1}{\epsilon \alpha^2}\right)}}
{\rm d}x
\nonumber\\
& \leq
\alpha^2 +
\frac{1}{\epsilon}
\int_{0}^{\infty}
\frac
{  1 }
{{\left( x + \frac{1}{\epsilon}\right)\left(x  + \frac{1}{\epsilon \alpha^2}\right)}}
{\rm d}x
 =
\alpha^2 +
\frac
{\alpha^2 \log\frac{1}{\alpha^2}}
{1 - \alpha^2}
\leq\frac{\alpha}{2} + \frac{\alpha^2 \left(\frac{1}{\alpha^2}\right)^{\frac{1}{2}}}{\frac{1}{2}\left(1 - \frac{1}{4}\right)} =C_5 \alpha,
\end{align}
where $C_5 = \frac{7}{8}$. The last inequality is because $\alpha \leq \frac{1}{2}$ and $\log x \leq 2 {x^{\frac{1}{2}}}$ for $x>0$.

The upper bound of  $I_{2, 2, 2}$ is derived as
\begin{align}
\label{I_1222}
I_{2, 2, 2}
& =
\frac{1}{\epsilon }
\int_{{\bar{\rho}}}^{\infty}
\frac{e^{-\frac{x}{P}}}{P \underbrace{(\epsilon x + 1)}_{\geq \epsilon x}}
\frac
{ 2 \alpha \bar{\rho} }
{\underbrace{\left(x A^2 + \frac{1}{\epsilon }\right)\left(x B^2 + \frac{1}{\epsilon }\right)}_{\geq \frac{1}{\epsilon^2}}}
{\rm d}x
+
\frac{1}{\epsilon }
\int_{{\bar{\rho}}}^{\infty}
\frac{\epsilon \overbrace{e^{-\frac{x}{P}}}^{\leq 1}}{\underbrace{(\epsilon x + 1)^2}_{\geq \epsilon^2 x^2}}
\frac
{ 2 \alpha \bar{\rho} }
{\underbrace{\left(x A^2 + \frac{1}{\epsilon }\right)\left(x B^2 + \frac{1}{\epsilon }\right)}_{\geq \frac{1}{\epsilon^2}}}
{\rm d}x
\nonumber\\
& \leq
\frac{1}{\epsilon }
\int_{{\bar{\rho}}}^{\infty}
\frac
{ e^{-\frac{x}{P}} \times 2 \alpha \bar{\rho} }
{P \times \frac{x}{\epsilon}}
{\rm d}x
+
\frac{1}{\epsilon }
\int_{{\bar{\rho}}}^{\infty}
\frac
{ \epsilon \times 2 \alpha \bar{\rho} }
{x^2 }
{\rm d}x
=
\frac{2 \alpha \bar{\rho}}{P}
\int_{{\bar{\rho}}}^{\infty}
\frac
{ e^{-\frac{x}{P}} }
{{ x}}
{\rm d}x
+
2 \alpha \bar{\rho}
\int_{{\bar{\rho}}}^{\infty}
\frac
{1}
{{x^2}}
{\rm d}x \nonumber\\
& =
\frac{2 \alpha \bar{\rho}}{P}
\int_{\frac{\bar{\rho}}{P}}^{\infty}
\frac
{ e^{-z} }
{{ z}}
{\rm d}z
+
2
\alpha
\leq \left[{2 e^{-\frac{\bar{\rho}}{P}}}\frac{ \log\left(1 + \frac{P}{\bar{\rho}}\right)}{\frac{P}{\bar{\rho}}}+2\right] \alpha \leq C_6 \alpha,
\end{align}
where $C_6 = 4$. After substituting \eqref{Final_I_1221} and \eqref{I_1222} into \eqref{I_122_OUT},  $I_{ 2, 2} \leq C_7 \alpha$, where $C_7 = C_5 + C_6$. Combined with \eqref{I_121_OUT}, \eqref{I_11_OUT}, \eqref{sum_I_2} and \eqref{I_11_OUT},  $I_{2} \leq C_8 \alpha$ and $\textmd{OUT}
 \left(\tilde{\textit{GQ}}_{{\sf mr}, {\sf it}}\right) - \textmd{OUT}_{{\sf mr}, {\sf it}}^{\sf opt} \leq 2(I_{1} + I_{2}) \leq C_9 \alpha$ when $M \geq 2$, where $C_8 = C_4 + C_7$ and $C_9 = 2 \left(C_3 + C_8\right)$. Letting $C_1 = \max\{C_2, C_9\}$,  $\textmd{OUT}
 \left(\tilde{\textit{GQ}}_{{\sf mr}, {\sf it}}\right) - \textmd{OUT}_{{\sf mr}, {\sf it}}^{\sf opt} \leq \frac{C_1}{M}$ for any $M \in \mathtt{N} - \{0\}$.

The upper bound on the average feedback rate of $\textit{DQ}_{{\sf mr}, {\sf it}}$ is derived as $\textmd{FR}\left(\textit{DQ}_{{\sf mr}, {\sf it}}\right)
   \leq 1 + 2\left\lceil \log_2\left(M + 1\right) \right\rceil
   \leq 2\log_2\left(M + 1\right) + 3$, which completes the proof. 
\end{IEEEproof}

\ifCLASSOPTIONcaptionsoff
  \newpage
\fi

%\bibliographystyle{IEEEtran}
%\bibliography{IEEEabrv,mybibfile}

\begin{thebibliography}{1}
\bibitem{Quantization_Interference}
K.~Anand, E.~Gunawan, and Y.~L.~Guan,
\newblock ``Beamformer design for the {MIMO} interference channels under limited channel feedback,''
\newblock \textit{IEEE Trans. Commun.}, vol. 61, no. 8, pp. 3246$-$3258, August 2013.
\bibitem{ErdemRelayFeedback}
E.~Koyuncu and H.~Jafarkhani,
\newblock ``Distributed beamforming in wireless multiuser relay-interference networks with quantized feedback,''
\newblock \textit{IEEE Trans. Inf. Theory}, vol. 58, no. 7, pp. 4538$-$4576, July 2012.
\bibitem{BDRaoTransmitBeamforming}
J.~C.~Roh and B.~D.~Rao,
\newblock ``Transmit beamforming in multiple-antenna systems with finite rate feedback: a VQ-based approach,''
\newblock \textit{IEEE Trans. Inf. Theory}, vol. 52, no. 3, pp. 1101$-$1112, Mar. 2006.
\bibitem{ErdemQuantization}
E.~Koyuncu and H.~Jafarkhani,
\newblock ``Very low-rate variable-length channel quantization for minimum outage probability,''
\newblock in \textit{IEEE Data Compression Conference (DCC)}, Mar. 2013, pp. 261$-$270.
\bibitem{Interference_Power_Control}
H.~Farhadi, C.~Wang, and M.~Skoglund,
\newblock ``Power control in wireless interference networks with limited feedback,''
\newblock in \textit{International Symposium on Wireless Communication Systems (ISWCS)}, Aug 2012, pp. 671$-$675.
\bibitem{Interference_Throughput}
H.~Farhadi, C.~Wang, and M.~Skoglund,
\newblock ``On the throughput of wireless interference networks with limited feedback,''
\newblock in \textit{IEEE International Symposium on Information Theory (ISIT)}, July 2011, pp. 762$-$766.
\bibitem{OptimalPowerControlSumRate}
A.~Gjendemsj{\o}, D.~Gesbert, G.~E.~{\O}ien and S.~G.~Kiani,
\newblock ``Optimal power allocation and scheduling for two-cell capacity maximization,''
\newblock in \textit{4th International Symposium on Modeling and Optimization in Mobile, Ad
Hoc and Wireless Networks}, 2006, pp. 1$-$6.
\bibitem{GLA}
Y.~Linde, A.~Buzo, and R.~M.~Gray,
\newblock ``An algorithm for vector quantizer design,''
\newblock in \textit{IEEE Trans. Commun.}, vol. 28, no. 1, pp. 84$-$95, 1980.
\bibitem{Handbook}
M.~Abramowitz and I.~A.~Stegun,
\newblock ``Handbook of mathematical functions,''
\newblock 1964.
\end{thebibliography}
\end{document}